%% file: main.tex
\documentclass[journal=jctcce,manuscript=article,layout=twocolumn]{achemso}

\input{preamble}
\author{Erik Verzijl}
\affiliation{Department of Chemistry and Pharmaceutical Sciences, Vrije Universiteit, De Boelelaan 1105, 1081 HV Amsterdam, The Netherlands}
\author{Arno F\"orster}
\email{a.t.l.foerster@vu.nl}
\affiliation{Department of Chemistry and Pharmaceutical Sciences, Vrije Universiteit, De Boelelaan 1105, 1081 HV Amsterdam, The Netherlands}

\title{Static Effective Hamiltonians for Molecular Systems through RPA-based Downfolding}

\begin{document}

\begin{abstract}
    Green's function-based downfolding methods construct effective Hamiltonians of reduced dimension that capture dynamical correlations of an electronic environment through effective potentials acting on the active space only. Using methods based on the constrained random phase approximation (cRPA) and moment RPA (mRPA), we construct static effective Hamiltonians that include screening through the environment. We derive expressions for the energy contribution from the environment and for the effective one- and two-body terms, taking into account double-counting corrections. cRPA requires additional consideration due to its frequency dependence, while mRPA provides a static Hamiltonian by construction. For the ground state energy of benzene and bond dissociation curves, we discuss the differences and similarities between the different flavors of RPA-based screening. We show that downfolding using cRPA describes both dynamical and strong correlation well, while mRPA and cRPA restricted to screening the particle-hole matrix elements can fail to describe bond dissociation due to a dominating dynamical correlation term. In the static limit, these two methods are shown to be almost indistinguishable.
\end{abstract}

\input{Chapters/1_Introduction}
\input{Chapters/2_Theory}
\input{Chapters/3_ComputationalDetails}
\input{Chapters/4_Results}
\input{Chapters/5_Conclusions}

\input{Chapters/appendix}


\section*{Data and Software Availability Statement}
The data underlying this study is available within the published article and its Supplementary Material. An example AMS input file and the custom script used to perform the subsequent NEVPT2 or high-level calculations are provided in the Supporting Information. Execution of the AMS input file requires a modified developer version of the code, which is accessible to licensed AMS developers or upon reasonable request. The script used to perform the subsequent NEVPT2 high-level calculation requires either the open-source PySCF software package\cite{Williams-Young2023AMethod} for FCI, or the open-source MACIS software package\cite{Sun2020RecentPackage} for ASCI.

\section*{Supplementary Material}
    Dissociation curves for \ce{N2} in cc-pVTZ basis, relative bonding energy curves, scripts used for the AMS and subsequent PySCF/MACIS calculation, and numerical data with: 1) Calculated screening per matrix element (two- out of eight-fold symmetry) for benzene in cc-pVDZ basis with (6,6) active space. 2) All calculated energies and $\text{Tr}[v^\text{eff}]$, both for benzene and the bond dissociation curves.
\section*{Author Declarations} The authors declare no conflict of interest.
\section*{Acknowledgments}
The authors thank Carlos Mejuto-Zaera for fruitful discussions. They also acknowledge use of the supercomputer facilities at SURFsara sponsored by NWO Physical Sciences, with financial support from the Netherlands Organization for Scientific Research (NWO). AF acknowledges funding through a VENI grant from NWO under grant agreement VI.Veni.232.013.

\bibliography{references.bib}

\begin{tocentry}
\includegraphics[width=\textwidth]{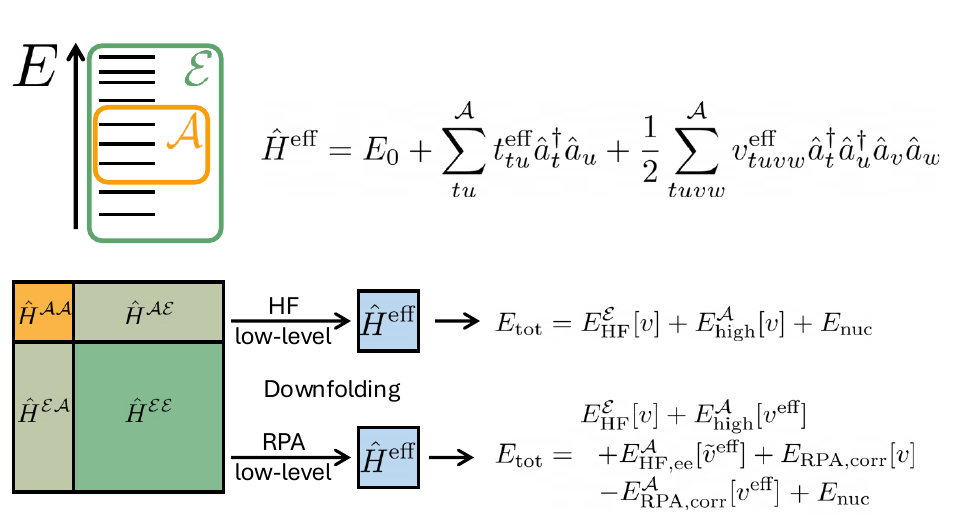}
\end{tocentry}

\end{document}

%% file: preamble.tex
\usepackage{graphicx}
\usepackage{subcaption}
\usepackage{pgfplots}
\pgfplotsset{compat=1.18}
\usepackage{xcolor}
\usepackage{multirow}
\usepackage[hidelinks]{hyperref}
\usepackage{braket}
\usepackage{amsmath}
\usepackage{amssymb}
\usepackage{bbold}
\usepackage{booktabs}
\usepackage[exponent-product = \cdot]{siunitx}
\newcommand{\otoprule}{\midrule[\heavyrulewidth]}
\usepackage{pdfpages}
\numberwithin{equation}{section}
\usepackage[version=3]{mhchem}
\usepackage{tikz}
\usepackage{adjustbox}
\usepackage{appendix}
\usepackage{bm}

\UseRawInputEncoding

\newcommand*{\dint}[1]{\ensuremath{\mathrm{d}{#1}}}
\newcommand*{\tb}[1]{\textbf{#1}}

\def\mA{\ensuremath{\mathcal{A}}}
\def\mR{\ensuremath{\mathcal{R}}}

\def\mE{\ensuremath{\mathcal{E}}}
\def\bigO{\ensuremath{\mathcal{O}}}

\def\ApB{\ensuremath{(\tb{A}+\tb{B})}}
\def\AmB{\ensuremath{(\tb{A}-\tb{B})}}
\def\XpY{\ensuremath{(\tb{X}+\tb{Y})}}
\def\XmY{\ensuremath{(\tb{X}-\tb{Y})}}
\def\ApmB{\ensuremath{(\tb{A}\pm\tb{B})}}
\newcommand*{\moment}[1]{\boldsymbol{\eta}^{(#1)}}
\newcommand{\mc}[1]{\multicolumn{1}{c}{#1}}


%% file: Chapters/1_Introduction.tex
\section{Introduction}

The quantitative description of strongly correlated systems presents a significant challenge in quantum chemistry. Only for small systems with small enough basis sets, the electronic Hamiltonian can be diagonalized exactly using full configuration interaction (FCI).\cite{Gao2024DistributedDeterminants} In many situations, selected configuration interaction (sCI) techniques~\cite{Huron1973IterativeWavefunctions, Harrison1991ApproximatingTheory, Tubman2020ModernMethod}, or the \emph{ab initio} density-matrix renormalization group (DMRG)~\cite{White1993Density-matrixGroups, Chan2002HighlyGroup, Chan2011TheChemistry, Olivares-Amaya2015ThePractice, Baiardi2020TheChallenges} can be used to obtain electronic ground states at near FCI accuracy,\cite{Sharma2014SpectroscopicDimer, Marie2024ReferenceTransitions, Zhai2026ClassicalImplications} with recent applications targeting impressively large active spaces of 89 electrons in 102 orbitals.\cite{Legeza2026HuntingTask} A novel FCI implementation reaches similar active space sizes\cite{Shayit2025NumericallyScale} leveraging an advanced lossless compression scheme.\cite{Hu2024SmallSupercomputer}

Nevertheless, a complete treatment of electron correlations for medium or even large molecules remains out of reach, and we must resort to approximate methods.
However, single-reference methods like coupled cluster with singles and doubles and perturbative triples (CCSD(T)) that scales as $\bigO(N^7)$\cite{Shavitt2009Many-bodyTheory,Raghavachari1989ATheories} completely break down in instances of strong electron correlation, where the single Slater determinant picture is invalid.\cite{Lyakh2012MultireferenceView, Henderson2014QuasiparticleInteractions, Bulik2015CanCorrelation}

To get the best of both worlds, quantum embedding methods divide the full system into regions that are treated with different levels of theory.\cite{Sun2016QuantumTheories} Here we focus on a single strongly correlated active region (\mA) defined by a subset of `active' orbitals obtained from a mean-field calculation, embedded in a weakly correlated environment (\mE). Many single-reference methods already make use of embedding, for instance, the frozen core approximation\cite{Roothaan1951NewTheory, Lykos1956OnRefinement} is technically an embedding method. Embedding methods designed to address multi-reference (MR) situations that are typically not referred to as one are complete active space (CAS) configuration interaction (CASCI),\cite{Levine2021CASOrbitals} where the active region is embedded in a HF mean-field or CAS-self-consistent field (CASSCF)\cite{Roos1980AApproach} that additionally optimizes the one-electron orbitals self-consistently. 

However, these embedding techniques do not describe the residual electron correlation outside the active space. These correlations are usually described using multi-reference (MR) perturbation theory (MRPT)\cite{Shavitt2009Many-bodyTheory} to low order, leading to methods such as complete active space second-order perturbation theory (CASPT2)\cite{Andersson1992Second-orderFunction, Ghigo2004ACASPT2} and n-electron valence state second-order perturbation theory (NEVPT2),\cite{Angeli2001IntroductionTheory, Angeli2002N-electronVariants} but a plethora of alternative methods of varying complexity have been designed for this purpose.\cite{Yanai2006CanonicalProblems,Neuscamman2010ATheory,Lyakh2012MultireferenceView,Evangelista2018Perspective:Correlation,LiManni2014MulticonfigurationTheory,Evangelista2014AProblems,Li2019MultireferenceGroup,Beran2021DensityConnection,Wang2025GeneralizedResummation} 

Evaluating these perturbative corrections is computationally expensive and limits the tractable active space size. Calculating the PT2 correlation in CASPT2 or NEVPT2 requires the costly calculation of four-particle reduced density matrices (RDM). Therefore, the PT2 step of a CASPT2 calculation begins to dominate computational timings already for active spaces of about 14 orbitals.\cite{Gaggioli2019BeyondCatalysis, Tubman2020ModernMethod} For the same reason, replacing the traditionally used FCI subroutines in the generation of the reference wave function by either DMRG\cite{Kurashige2011Second-orderDimer, Guo2016N-ElectronPolyp-Phenylenevinylene} or sCI\cite{Park2023DynamicASCI-DSRG-MRPT2} does not automatically allow for treating the same large active spaces that these methods can normally handle. Despite significant progress,\cite{Kurashige2011Second-orderDimer, Guo2016N-ElectronPolyp-Phenylenevinylene, Cheng2022Post-DensitySpaces} they are currently limited to active spaces of around 40 orbitals.\cite{Park2023DynamicASCI-DSRG-MRPT2, Cheng2022Post-DensitySpaces}

For this reason, it is desirable to develop alternative approaches that can provide a balanced description of static and dynamical correlations without performing MRPT calculations. Over the years, numerous approaches have emerged that avoid this separation altogether, instead treating the coupling of the active region to its environment iteratively rather than sequentially. Notable examples are density functional embedding theory (DFET),\cite{Cortona1991Self-consistentlyCalculations, Wesolowski1993FrozenMolecules, Manby2012AScheme, Libisch2014EmbeddedApplications, Tamukong2017AccurateOrthogonality} density matrix embedding theory (DMET),\cite{Knizia2012DensityTheory, Knizia2013DensityTheory} and the Green's function-based dynamical mean-field (DMFT),\cite{Georges1996DynamicalDimensions, Lin2011DynamicalChemistry, Zgid2011DynamicalPerspective, Turkowski2012DynamicalNanostructures, Chibani2016Self-consistentConcept} and self-energy embedding theories (SEET)\cite{Kananenka2015SystematicallySystems, Lan2015Communication:Chemistry, Lan2017GeneralizedTheory}. The philosophy behind these methods is markedly different from that of MRPT. Instead of constructing a wave function expansion, they self-consistently construct effective Hamiltonians or potentials that describe the hybridization of active electrons with their environment. 

In this work, we design effective Hamiltonians through downfolding. These Hamiltonians explicitly include the residual, mostly dynamical correlation of the environment through an effective potential acting on the active space only, and can be diagonalized using standard solver like FCI, DMRG, or sCI. In contrast to the aforementioned embedding methods, here dynamical and static correlations are treated sequentially (in this work we refer to the residual environment correlations as dynamical, and to the ones between the active space electrons as static), but in contrast to MRPT-based approaches, the dynamical correlations are treated first. Therefore, the calculations of perturbative corrections as in CASPT2 and NEVPT2, along with the common difficulties associated with these MRPT approaches, are avoided. Instead, the complexity of the problem is transferred to the construction of the effective Hamiltonian.

We use many-body perturbation theory (MBPT),\cite{Fetter2003QuantumSystems, Martin2016InteractingApproaches} within the random phase approximation (RPA),\cite{Macke1950UberElektronenkondensation, Bohm1953AGas} commonly used in combination with the $GW$ approximation,\cite{Hedin1965NewProblem, Reining2018TheLimitations, Marie2024ThePerspective} to renormalize the two-body interactions of the active space Hamiltonian.
In condensed matter physics, the constrained RPA (cRPA)\cite{Aryasetiawan2004Frequency-dependentCalculations, Aryasetiawan2006CalculationsFirst-principles} has emerged as a widespread method to calculate effective two-body interactions to study for instance quantum defects\cite{Bockstedte2018AbBits, Ma2021QuantumMaterials, Huang2022SimulatingComputers, Sheng2022GreensTheory, Romanova2023DynamicalStates, Chen2025AdvancesTheory} or superconductivity,\cite{Hirayama2017Low-energyTheory, Hirayama2018AbSuperconductors, Hirayama2019EffectiveRenormalization, Nilsson2019DynamicallyExceptions, Ohgoe2020AbSuperconductor} often combined with the constrained $GW$ (c$GW$) approximation.\cite{Werner2015DynamicalLa2CuO4, Hirayama2017Low-energyTheory, Nakamura2021RESPACK:Material, Romanova2023DynamicalStates} These methods provide a favorable balance between computational cost and accuracy and allow for applications to systems with several thousand electrons,~\cite{Vlcek2018SwiftCompression, Spadetto2023TowardAccuracy, Yeh2023Low-ScalingSampling} while also capturing the dynamical screening that is believed to be the dominant correlation channel for weakly correlated environments.\cite{Reining2018TheLimitations, Shinaoka2015AccuracyApproximation, vanLoon2021RandomApproximation}

On the other hand, the accuracy of these methods has been questioned, with the cRPA often overscreening the interactions.\cite{Honerkamp2018LimitationsDownfolding, vanLoon2021RandomApproximation} Furthermore, the fact that Green's function-based downfolding leads to frequency-dependent interactions poses an additional difficulty. Although specialized solvers\cite{Gull2011Continuous-timeModels} can deal with frequency-dependent interactions, the solver typically employed in quantum chemistry cannot. One is therefore forced to diagonalize the effective Hamiltonian at judiciously chosen frequencies.\cite{Romanova2023DynamicalStates} In practice, a fixed value of $\omega$, often $\omega = 0$ is typically used,\cite{Sheng2022GreensTheory} even though this is in principle an uncontrolled approximation. 

Avoiding the difficulties with frequency-dependent interactions, Scott and Booth recently introduced the moment RPA (mRPA)\cite{Scott2024RigorousSystems} theory that is based on the density-density (dd) response moments.\cite{Scott2021ExtendingTheory, Scott2023ATheory} 
Additionally, mRPA avoids a strict partitioning constraint of cRPA, which requires the absence of mean-field coupling between the active space and environment so that the polarizability can be separated without introducing double-counting errors. This is primarily an advantage for solids, where the active bands are often entangled with the environment and cannot easily be isolated.\cite{Kaltak2025ConstrainedMethod} For molecules, the use of a canonical orbital basis ensures that there is no coupling between the active space and the environment by construction, so the constraint is always satisfied.

cRPA and mRPA have been compared recently,\cite{Pauli2025StaticModel} but not in a molecular context. Generally, applications of RPA-downfolded Hamiltonians to molecules are rare. We are only aware of the work by Dvorak et al.\cite{Dvorak2019DynamicalFunctions,Dvorak2019QuantumGW/BSE} who used cRPA-downfolded Hamiltonians to calculate excited states of small molecules with promising results. In this paper, we report implementations of cRPA and mRPA and benchmark them in the context of molecules. Inspired by the mRPA approach,\cite{Scott2024RigorousSystems} we will also assess a variant of cRPA with screening restricted to the particle-hole matrix elements only.
We will then discuss the different screening methods together with the partitioning and treatment of the active region and the environment, derive expressions for the one- and two-body terms of the effective Hamiltonian, and assess their suitability for calculating ground state energies.
We focus primarily on bond dissociation as prototypical MR problems. We investigate the magnitude of screening of the individual two-body integrals and the accuracy of the resulting effective Hamiltonians on total energies, both for geometries and dissociation curves.

%% file: Chapters/2_Theory.tex
\section{Theory}

We aim at constructing effective Hamiltonians $\hat{H}^\text{eff}$, which act only in an active space,
\begin{equation}\label{eq:Heff}
    \hat{H}^\text{eff}=E_0+\sum_{tu}^\mA t_{tu}^\text{eff}\hat{a}_t^\dagger\hat{a}_u+\frac{1}{2}\sum_{tuvw}^\mA v_{tuvw}^\text{eff}\hat{a}_t^\dagger\hat{a}_u^\dagger\hat{a}_v\hat{a}_w \;.
\end{equation}
Here, indices $t$, $u$, $v$, and $w$ are consistently used as general active spin-orbitals. The occupied and virtual spin-orbitals will be referred to with indices $i$, $j$ and $a$, $b$, respectively. General spin-orbitals have indices $p$, $q$, $r$, and $s$. Two-electron integrals will be denoted as 
\begin{equation}
\begin{aligned}
   & v_{pq,rs} = \\
   &\int \int \dint{\bm{r}}\dint{\bm{r}'} 
   \phi^*_{p}(\bm{r}) 
   \phi_{q}(\bm{r})
   v(\bm{r},\bm{r}')
   \phi^*_{r}(\bm{r}') 
   \phi_{s}(\bm{r}')  \;,
   \end{aligned}
\end{equation}
where $\phi_{p}$ is a single-particle wave function and $v(\bm{r},\bm{r}')$ some (potentially effective) Coulomb interaction. This notation corresponds to the chemist notation $(pq|rs)$. In Eq.~\eqref{eq:Heff}, the environment orbitals are integrated out so that the interaction is downfolded onto the effective one- and two-body terms. Therefore, in addition to the nuclear repulsion term, the zero-body term $E_0$ also includes the contribution of these eliminated degrees of freedom to the total energy:
\begin{equation}\label{eq:E0}
    E_0 = E_{\text{nuc}} + E^{\mE} \;.
\end{equation}
The CASCI Hamiltonian is one of the simplest downfolded Hamiltonians. The effective two-body terms are simply left bare, $v^\text{eff}=v$, and the environment enters the effective Hamiltonian through the one-body terms, as a bare interaction with the mean-field.\cite{Yalouz2021AStates} This gives
\begin{equation}\label{eq:Hcasci}
\begin{aligned}
    \hat{H}^\text{eff}_\text{CASCI}=& E_0+\sum_{tu}^\mA f_{tu}\hat{a}_t^\dagger\hat{a}_u \\
    &+\frac12\sum_{tuvw}^\mA v_{tuvw}\hat{a}_t^\dagger\hat{a}_u^\dagger\hat{a}_v\hat{a}_w \;.
\end{aligned}
\end{equation}
where
\begin{equation}\label{eq:HcasciOneBody}
    f_{tu}=h^{(0)}_{tu}+\sum_i^\mE\left(v_{tu,ii}-v_{ti,iu}\right)\rho_{ii} \;,
\end{equation}
and $E^\mE$ only contributes to $E_0$ through a HF-term,
\begin{equation}\label{eq:EnvEhf}
\begin{aligned}
    E^\mE=&E^\mE_\text{HF,core}+E^\mE_\text{HF,ee}\\
    =&\sum_i^\mE h^{(0)}_{ii}\rho_{ii}+\frac12\sum_{ij}^\mE\left(v_{ii,jj}-v_{ij,ji}\right)\rho_{ii}\rho_{jj} \;.
\end{aligned}
\end{equation}
Solving this effective CASCI Hamiltonian accounts for all electronic interactions within the active space. However, the environment is only accounted for as a static mean-field. 

In addition, we also want to fold the dynamical electron correlations onto the effective Hamiltonian. We will use the RPA, defined in the space of particle-hole (de)excitations from a single reference state\cite{Chong1995RecentMethods,Furche2001OnTheory,Bruneval2016MolgwClusters}
\begin{equation}\label{eq:RPAcasida}
    \begin{pmatrix}\tb{A}&\tb{B}\\-\tb{B}&-\tb{A}\end{pmatrix}\begin{pmatrix}\tb{X}_s\\\tb{Y}_s\end{pmatrix}=\begin{pmatrix}\tb{X}_s\\\tb{Y}_s\end{pmatrix}\boldsymbol{\Omega}_\nu \;,
\end{equation}
where $\tb{A}$ and $\tb{B}$ are defined as 
\begin{subequations}
\begin{align}
    \tb{A}_{ia}^{jb}&=-v_{ia,jb}+\left(\epsilon_a-\epsilon_i\right)\delta_{ij}\delta_{ab}\label{eq:A}\\
    \tb{B}_{ia}^{jb}&=-v_{ia,jb}\label{eq:B}\;,
\end{align}
\end{subequations}
as in Ref.~\citenum{Bruneval2016MolgwClusters}. In practice, only the matrices $\left(\tb{A}-\tb{B}\right)^{1/2}$ and $\left(\tb{A}+\tb{B}\right)$ are constructed, using the block structure of \eqref{eq:RPAcasida}. This gives the recast eigenvalue problem
\begin{equation}\label{eq:RPArecast}
    \left(\tb{A}-\tb{B}\right)^\frac12\left(\tb{A}+\tb{B}\right)\left(\tb{A}-\tb{B}\right)^\frac12\tb{Z}=\tb{Z}\boldsymbol\Omega^2 \;.
\end{equation}
From this, the left and right eigenvectors can be obtained, using
\begin{subequations}
\begin{align}
    \tb{X}&=\frac12\left[\boldsymbol\Omega^{-\frac12}\left(\tb{A}-\tb{B}\right)^\frac12+\boldsymbol\Omega^\frac12\left(\tb{A}-\tb{B}\right)^{-\frac12}\right]\tb{Z}\\
    \tb{Y}&=\frac12\left[\boldsymbol\Omega^{-\frac12}\left(\tb{A}-\tb{B}\right)^\frac12-\boldsymbol\Omega^\frac12\left(\tb{A}-\tb{B}\right)^{-\frac12}\right]\tb{Z} \;,
\end{align}
\end{subequations}
with the normalization condition
\begin{equation}\label{eq:XYnormalization}
    \begin{pmatrix}\tb{X}&-\tb{Y}\\-\tb{Y}&\tb{X}\end{pmatrix}^T\begin{pmatrix}\tb{X}&\tb{Y}\\\tb{Y}&\tb{X}\end{pmatrix}=\tb{I} \;.
\end{equation}
From neutral excitations $\boldsymbol{\Omega}$ and transition densities $\left(\tb{X}+\tb{Y}\right)$, the screened interaction $W$ is calculated via
\begin{equation}\label{eq:W}
\begin{aligned}
    W_{pq}^{rs}&(\omega)=v_{pq,rs}+\sum_\nu w^\nu_{pq}w^\nu_{rs}\\&\times\left(\frac{1}{\omega-\Omega_\nu+i\eta^+}-\frac{1}{\omega+\Omega_\nu-i\eta^+}\right)
\end{aligned}
\end{equation}
with
\begin{equation}\label{eq:w}
    w^\nu_{pq}=\sum_{ia}v_{pq,ia}\left[\left(\tb{X}+\tb{Y}\right)\right]^\nu_{ia} \;.
\end{equation}
Eq~\eqref{eq:W} gives the full screening of the electron-electron interaction through all particle-hole excitations in the system. Therefore, using $v^\text{eff}=W(\omega)$ would introduce double counting, as the screening effects of interactions within the active space are already implicitly accounted for by diagonalizing the active space Hamiltonian. 

In the cRPA,\cite{Aryasetiawan2004Frequency-dependentCalculations, Aryasetiawan2006CalculationsFirst-principles} this double-counting is avoided by removing active-active excitations from $\tb{A}$ and $\tb{B}$, so that reduced eigenvectors $\left(\tb{X}+\tb{Y}\right)^\mR$ and eigenvalues $\boldsymbol{\Omega}^\mR$ are obtained by solving the eigenvalue problem in eq~\eqref{eq:RPAcasida} with reduced matrices $(\tb{A}\pm\tb{B})^\mR$,
\begin{equation}
    \left(\tb{A}\pm\tb{B}\right)^\mR_{ia,jb}=\begin{cases}0,&i,a,j,b\in\mA\;,\\\left(\tb{A}\pm\tb{B}\right)_{ia,jb}, & \text{otherwise} \;,
\end{cases}
\end{equation}
Because active-active excitations are projected out, active-active neutral excitations will be zero and have null eigenvectors, and therefore do not contribute to the screening. To obtain a static effective Hamiltonian, we need to eliminate the frequency dependence in eq~\eqref{eq:W}. Often, the static limit $\omega=0$  corresponding to the long-time limit is taken, which is equivalent to the approximation that the environment reacts (and relaxes) instantaneously to fluctuations of the active space electrons. For strongly correlated systems, where the energy levels are not clearly separated, this is an unjustified approximation, and the screening frequency requires additional consideration.\cite{Canestraight2025RenormalizationDownfolding} Different approaches include: (1) taking a fixed $\omega$ for all screened matrix elements (can be the static limit) (2) using a mono- or multi-pole model\cite{Chang2024DownfoldingApproach} (3) element-wise using quasi-particle energies\cite{Canestraight2025RenormalizationDownfolding}. In this work, we only consider (1), in future work (2) and (3) will also be considered. Due to the pole-structure of $W$, we have a similar restriction on the allowed frequency-range as described by Van Loon~et al, \cite{vanLoon2021RandomApproximation} where $\omega$ must be smaller than the mean-field gap $E_\text{KS}$. In our case, this corresponds to the restriction $\omega<\min\{\boldsymbol{\Omega}^\mR\}$.

mRPA is designed to eliminate the dependence on the screening frequency altogether,\cite{Scott2021ExtendingTheory,Scott2023ATheory} so that when applied to downfolding, it gives a static effective interaction.\cite{Scott2024RigorousSystems} It does so through the evaluation of the moment expansion of the dd-response,
\begin{equation}
\label{eq:eta_n}
    \moment{n}=-\frac{1}{\pi}\int_0^\infty\Im[\boldsymbol{\eta}(\omega)]\omega^n\dint{\omega},
\end{equation}
which can be used as a constraint to be matched up to a desired order. Evaluating the frequency integral gives a simple expression for each $n^\text{th}$ order dd-moment:
\begin{equation}
    \moment{n}=\left(\tb{X}+\tb{Y}\right)\boldsymbol{\Omega}^n\left(\tb{X}+\tb{Y}\right)^T \;.
\end{equation}
In Appendix~\ref{sec:A_mRPA}, we show some relations between dd-moments, where, most importantly, the zeroth- and first-order dd-moments are related by
\begin{equation}\label{eq:eta1}
    \moment{1}=\left(\tb{A}-\tb{B}\right)=\moment{0}\left(\tb{A}+\tb{B}\right)\moment{0} \;,
\end{equation}
and we can express the bare Coulomb interaction in terms of these moments:
\begin{equation}\label{eq:vEffmRPA}
    v=\frac{1}{2}\left(\left(\moment{0}\right)^{-1}\moment{1}\left(\moment{0}\right)^{-1}-\moment{1}\right) \;.
\end{equation}
Now, following Scott and Booth, instead of using $\moment{0}$ and $\moment{1}$ for the full system, we project these quantities on the active space, so that a static effective interaction that lives in the active space is obtained.\cite{Scott2024RigorousSystems} This static effective interaction conserves by construction the zeroth- and first-order dd-moments.\cite{Scott2024RigorousSystems}

Although both mRPA and cRPA are based on the same low-level method, the resulting $v^\text{eff}$ are very different. mRPA conserves $\moment{0}$ and $\moment{1}$, whereas cRPA only conserves $\moment{1}$ in the static limit, $\omega\rightarrow0$.\cite{Scott2024RigorousSystems} Furthermore, because $\left(\tb{X}+\tb{Y}\right)^\mR$ and $\boldsymbol{\Omega}^\mR$ are in the basis of particle-hole excitations, mRPA only screens elements $v_{ia,jb}$, while \eqref{eq:W} shows that cRPA screens all elements $v_{pq,rs}$. To connect the two methods, we also propose a screening method that we call cRPAph, which can be understood as the combination of the two methods. We perform a cRPA calculation but, as in mRPA, we only screen the particle-hole matrix elements $v_{ia,jb}$.

If we would simply plug $v^\text{eff}$ in the effective Hamiltonian for the two-body terms, we would neglect the fact that the Hartree and exchange-correlation interactions are also screened, leading to double counting. We therefore introduce the double-counting term, 
\begin{equation}\label{eq:tDC}
    t^\text{DC}_{tu}=\sum^\mA_{vw}\left(\Tilde{v}^\text{eff}_{tu,vw}-\Tilde{v}_{tw,vu}^\text{eff}\right)\rho_{vw} \;,
\end{equation}
with $\Tilde{v}^\text{eff}=v^\text{eff}-v$, resulting in the effective Hamiltonian
\begin{equation}\label{eq:H_eff_rpa}
\begin{aligned}
    \hat{H}^\text{eff}=E_0+&\sum_{tu}^\mA\left(f_{tu}-t_{tu}^\text{DC} \right)\hat{a}_t^\dagger\hat{a}_u\\&+\frac12\sum_{tuvw}^\mA v^\text{eff}_{tuvw}\hat{a}_t^\dagger\hat{a}_u^\dagger\hat{a}_v\hat{a}_w \;,
\end{aligned}
\end{equation}
For cRPAph and mRPA, where only particle-hole matrix elements are screened, we only need to account for double counting in the exchange-like term in Eq.~\eqref{eq:tDC} where $v$ and $w$ are occupied orbitals and $t$ and $u$ are virtual orbitals, since all other terms vanish.
In the appendix~\ref{sec::appendix}, we present a rigorous derivation of this Hamiltonian for the cRPA case, starting from the Baym--Kadanoff $\Phi$-functional in the $GW$ approximation.\cite{Baym1961ConservationFunctions, Almbladh1999VariationalTheories} As cRPAph uses the same framework as cRPA, this derivation also strictly holds for the former. mRPA uses a different framework, but, due to the similarities with cRPAph that we will discuss later, we can pragmatically assume the same form here as well. The double-counting correction has already been discussed previously,\cite{Pfaffle2021ScreenedCenters,Ma2021QuantumMaterials,Sheng2022GreensTheory} and Eq.\eqref{eq:tDC} is identical to the ones used in Refs.~\citenum{Pfaffle2021ScreenedCenters} and \citenum{Ma2021QuantumMaterials}. 
In those works, the double counting correction was stated to be exact from a HF starting point, but only approximate within a DFT formulation. However, there, $t^\text{eff}$ is constructed by subtracting Hartree- and exchange-like interactions from the active space from $H^\text{KS}$, thereby not exactly removing all double counted exchange-correlation from $H^\text{KS}$. In this work, we construct $t^\text{eff}$ explicitly as $f^\mE-t^\text{DC}$, so no correlation effects are double counted, and the formulation is exact no matter the starting point. 

As stated in the introduction, we show the effect of the different screening methods by evaluating the total ground state energies. Because we use RPA as the low-level method, $E^\mE$ in Eq.~\eqref{eq:E0} also contains dynamical correlation from the integrated out environment orbitals, which, as desired, is calculated simultaneously and without MRPT calculations. In Appendix~\ref{sec:A_E0} we derive an expression for $E^\mE$, which includes both the RPA correlation energy of the environment, and corrects for the mean-field energy of the active space:
\begin{equation}
\label{eq:environment_correlation}
\begin{aligned}
    E^{\mE} = & E^\mE_\text{HF,core} + E^{\mE}_{\text{HF,ee}} + E^{\mA}_{\text{HF,ee}}[\tilde{v}^{\text{eff}}] \\
    &+ 
    E_{\text{RPA, corr}}[v] - E^{\mA}_{\text{RPA, corr}}[v^{\text{eff}}] \;.
\end{aligned}
\end{equation}
The correction term $E^{\mA}_{\text{HF,ee}}[\tilde{v}^{\text{eff}}]$ is only non-zero for cRPA, since for mRPA and cRPAph, the mean-field energy contribution of the active space is conserved. It can be interpreted as a residual correlation term that arises from the non-separability of the logarithmic $GW$-$\Phi$-functional\cite{Almbladh1999VariationalTheories} under the trace. It would not arise for a $\Phi$-functional that is a polynomial in the Coulomb interaction.
The RPA correlation energy is calculated with the Klein functional,\cite{Furche2008DevelopingModel}
\begin{equation}
\begin{aligned}
    E_\text{RPA,corr}=&\frac12\sum_\nu\Omega_\nu-\frac{1}{2}\sum_{ia}\left[\epsilon_a-\epsilon_i\right]\\&-\sum_{ia}v_{ia,ia} \;.
\end{aligned}
\end{equation}
$E_\text{RPA,corr}^\mA[v^\text{eff}]$ is calculated by solving \eqref{eq:RPAcasida} with screened interaction and active orbitals only. Equivalently, when computed from a mRPA perspective, projection on the active space in Eq.~\eqref{eq:mRPA_EcorrA} yields identical energies. 

The use of $v^\text{eff}$ in Eq.~\eqref{eq:H_eff_rpa} explicitly describes how the interactions between active space electrons are screened through particle-hole excitations in the environment. Therefore, this defines a polarizable embedding scheme that distinguishes it from mechanical embedding approaches that evaluate $\hat{H}^\text{eff}$ with a bare interaction among active orbitals. Mechanical embedding has been applied in heterogeneous catalysis\cite{Boese2013AccurateCOMgO001, Alessio2019ChemicallySurface} and to study iron-sulfur clusters.\cite{Zhai2023MultireferenceClusters} We will show that the polarizable embedding approach defined by Eq.~\eqref{eq:H_eff_rpa} yields certain improvements over mechanical embedding, even for small molecules.

%% file: Chapters/3_ComputationalDetails.tex
\section{Computational details}
We calculated all effective Hamiltonians using the BAND\cite{teVelde1991PreciseStructures} engine of a modified development version of the Amsterdam Modeling Suite (AMS2025.206)\cite{Baerends2025TheSuite} and stored them in the form of FCIDUMP files,\cite{Knowles1989AProgram} containing all necessary one- and two-electron integrals. In this work, all active spaces for which we constructed downfolded Hamiltonians were chosen small enough so that FCI could be used as the high-level method, in which case PySCF's\cite{Sun2020RecentPackage} FCI-solver was used to obtain $E_\text{FCI}^\mA$. For consistency, the NEVPT2 and high-level reference calculations also used FCIDUMP files as direct input. NEVPT2 reference energies were obtained using PySCF's MRPT module. High-level reference energies were calculated either via exact diagonalization with PySCF's FCI solver or, for larger systems where FCI is intractable, using adaptive sampling configuration interaction (ASCI)\cite{Tubman2016AMethod,Tubman2020ModernMethod} as implemented in MACIS.\cite{Williams-Young2023AMethod} Example scripts are provided in the Supplemental Information.


All calculations were performed with restricted spin, with all default AMS settings, unless specifically stated otherwise. We used Hartree--Fock or PBE\cite{Perdew1996GeneralizedSimple} for the prior mean-field calculation. The implementation in AMS allows for both the use of Slater-type orbitals (STO)\cite{VanLenthe2003Optimized1118} and Gaussian-type orbitals (GTO)\cite{Dunning1971GaussianAtoms}, taken from Basis Set Exchange.\cite{Pritchard2019NewCommunity} The basis set used will be explicitly mentioned when discussing the results.

Active spaces will be denoted as $(n,m)$, where $n$ is the number of active electrons and $m$ is the number of active orbitals. If not obvious, it will be accompanied by a specification of which orbitals comprise the active space.

%% file: Chapters/4_Results.tex
\section{Results}\label{sec:Results}
\subsection{Benzene - frequency dependence}\label{sec:BenzeneOmega}

\begin{figure*}[hbt!]
    \centering
    \input{Images_new/1_BenzeneHeatmaps}
    \caption{$-|v^\text{eff}(\omega=0)-v|$ with a) cRPA,  b) mRPA, and c) cRPAph for benzene in cc-pVDZ basis with a (6,6) active space. Each cell shows an element $-|v_{tu,vw}^\text{eff}(\omega=0)-v_{tu,vw}|$, on the $x$-axis all combinations $t,u$ with $u\geq t$ and on the $y$-axis all combinations $v,w$ with $w\geq v$. The sign does not change upon screening, so absolute values are used to show the magnitude of screening. To emphasize that screening lowers the magnitude for each matrix element, we plot this difference as negative values.}
    \label{fig:1_BenzeneHeatmaps}
\end{figure*}
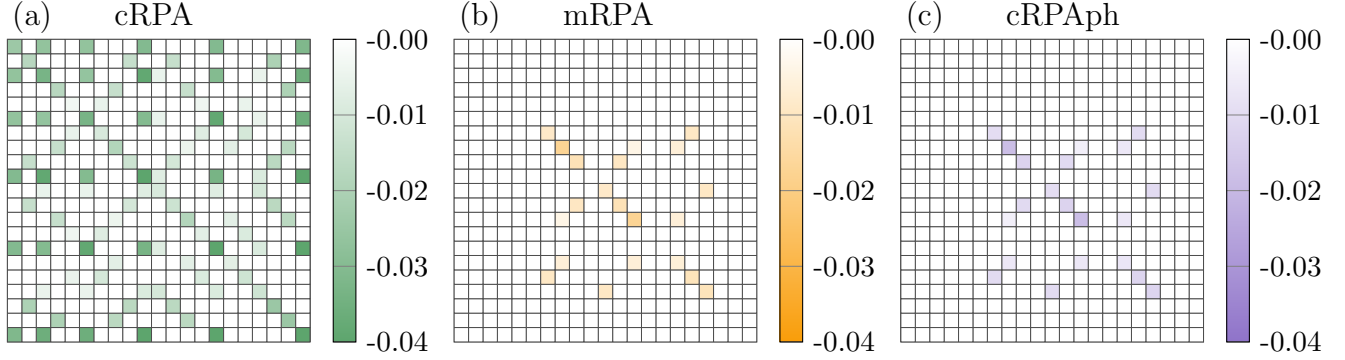

To show that the screening methods described above are applicable as downfolding methods for molecules, we report some examples. First, we present results for benzene in cc-pVDZ basis, for which the ground state energy has been well-studied.\cite{Eriksen2020TheBenzene} Before considering the effect of the screening frequency, Fig.~\ref{fig:1_BenzeneHeatmaps} shows the screening per matrix element for the different screening methods at the zero-frequency limit. Here we consider an (6,6) active space, containing all six linear combinations of $2p_z$ orbitals. cRPA and mRPA mainly differ in the matrix elements that are screened. As described by Scott and Booth, mRPA only screens the particle-hole matrix elements, whereas cRPA screens all matrix elements.\cite{Scott2024RigorousSystems} The reason mRPA only screens the particle-hole channel comes directly from its construction, namely from the spectral moments that are determined by the eigenvectors and eigenvalues of Eq.~\eqref{eq:RPArecast}, which are obtained from an eigenproblem in particle-hole basis. cRPA, on the contrary, uses Eq.~\eqref{eq:W}, and therefore all matrix elements are screened. If we limit the cRPA to screening the particle-hole matrix elements only, we see in Fig~\ref{fig:1_BenzeneHeatmaps}c that, at the zero-frequency limit, the screening is almost identical to mRPA.

\begin{figure}[hbt!]
    \centering
    \includegraphics[width=\linewidth]{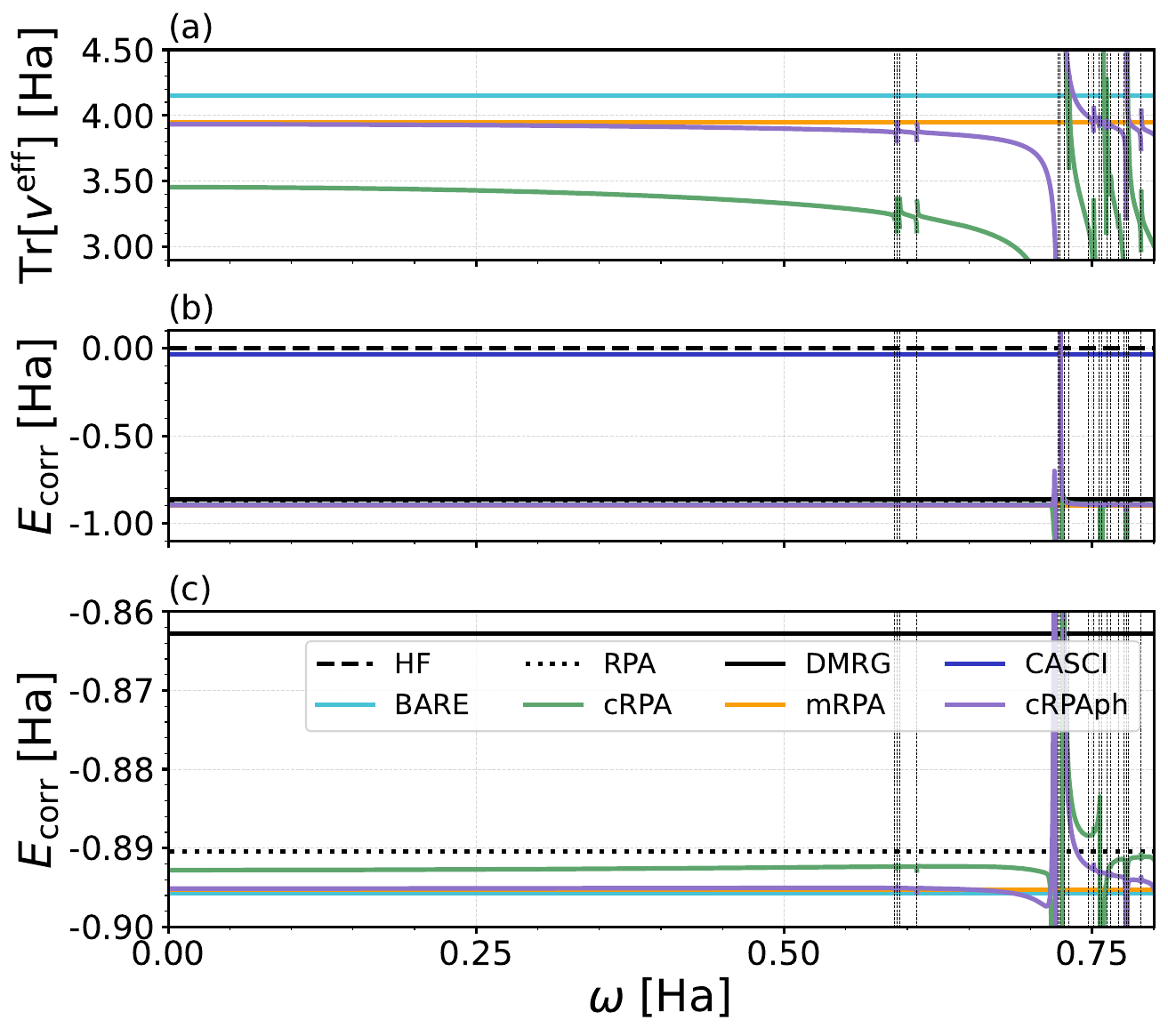}
    \caption{Benzene in cc-pVDZ basis with (10,10) active space. a) $\text{Tr}[v^\text{eff}]$, b) $E_\text{corr}$, c) $E_\text{corr}$ zoomed in at the DMRG reference and the RPA-based downfolding methods. The dashed vertical lines are the values of $\Omega_\nu\in\boldsymbol{\Omega}^\mR$. We used $\eta^+=10^{-4}$ as broadening factor.}
    \label{fig:2_BenzeneOmega}
\end{figure}

We will now further compare the different screening methods, also taking into account the dependence on the screening frequency for cRPA and cRPAph. To first demonstrate the correctness of the implementation, we now consider the same (10,10) active space (HOMO-4 to LUMO+4) as used by Scott and Booth.\cite{Scott2024RigorousSystems} 
Comparing Fig.~\ref{fig:2_BenzeneOmega}a, that the trace of $v^\text{eff}$, with Fig.~2 in Ref.~\citenum{Scott2024RigorousSystems}, we clearly reproduce their results with bare and mRPA screening, as well as with cRPA screening for $\omega<\min(\boldsymbol{\Omega}^\mR)$. For larger screening frequencies, we enter the domain where all the poles of the RPA response function reside. This we can clearly see in Fig.~\ref{fig:2_BenzeneOmega}a, but the visibility of the individual poles depends on the number of points and the value used for $\eta^+$ when evaluating Eq.~\eqref{eq:W}. Especially in the domain where the eigenvalues $\boldsymbol{\Omega}^\mR$ are closely together, it becomes very erratic. Additionally, if one were to analytically continue from a grid of imaginary frequencies, the pole-like structure would naturally not emerge. This supports the restriction $\omega<\min\{\boldsymbol{\Omega}^\mR\}$. In Fig.~\ref{fig:2_BenzeneOmega}a, we also see that the $\text{Tr}[v^\text{eff}]$ obtained with mRPA and cRPAph are almost identical for $\omega=0$, and only differ at a higher screening frequency. 

The differences and similarities between the screening methods are also seen when evaluating the correlation energies (Fig.~\ref{fig:2_BenzeneOmega}b/c). Here, for reference, we also include $E_\text{CASCI}$, obtained from Eqs.~\eqref{eq:Hcasci}-\eqref{eq:EnvEhf}, and the near-exact correlation energy calculated with DMRG.\cite{Eriksen2020TheBenzene} Since, in its equilibrium geometry, benzene has one dominant determinant and we have a small active space, CASCI does not capture much correlation and is very close in energy to HF. The RPA-based downfolding methods are almost identical to plain RPA. Because the active space contains few orbitals, almost all excitations are included in $\boldsymbol{\Omega}^\mR$, so the RPA correlation from the environment is very close to the RPA correlation from the full system. As we see, the correlation energy obtained from solving the effective Hamiltonian is (only slightly) bigger than this difference, resulting in a slightly higher correlation energy compared to plain RPA. If we leave the effective two-body terms unscreened, the active orbitals are naturally correlated the most. Screening reduces the active space correlation energy, which we see if we compare mRPA-$v^{\text{eff}}$ to an unscreened interaction $v^{\text{eff}} = v$. Since mRPA is almost identical to cRPAph at the zero-frequency limit (Fig.~\ref{fig:2_BenzeneOmega}a and Fig~\ref{fig:1_BenzeneHeatmaps}), these methods yield almost identical correlation energies. Screening within cRPA reduces the correlation energy further. Fig.~\ref{fig:2_BenzeneOmega}b/c also shows that, for this system with this active space, the correlation energy is insensitive to the screening frequency. Only once we reach the domain where the restriction $\omega<\text{min}(\boldsymbol{\Omega}^\mR)$ is not satisfied anymore, the energies are, as expected, erratic.

\subsection{Dissociation}
\begin{table*}[hbt!]
    \small
    \centering
    \caption{Equilibrium bond length in \AA}
    \begin{tabular}{l|*{9}{S[table-format=1.3, table-column-width=1.1cm]}}
    \toprule
     & \mc{FCI/ASCI} & \mc{HF} & \mc{RPA} & \mc{CASCI} & \mc{NEVPT2} & \mc{BARE} & \mc{cRPA} & \mc{mRPA} & \mc{cRPAph} \\
    \otoprule
    \ce{H2} (HF)  & 0.743 & 0.734 & 0.735 & 0.740 & 0.741 & 0.734 & 0.734 & 0.734 & 0.734 \\
    \ce{H2} (PBE) & 0.743 & 0.734 & 0.744 & 0.747 & 0.747 & 0.735 & 0.738 & 0.738 & 0.738 \\
    \ce{H6} (4,4) & 0.996 & 0.990 & 0.989 & 1.001 & 0.998 & 0.993 & 0.992 & 0.993 & 0.993 \\
    \ce{H6} (6,6) & 0.996 & 0.990 & 0.989 & 1.020 & 1.000 & 0.992 & 0.992 & 0.993 & 0.993 \\
    \ce{N2}       & 1.120 & 1.080 & 1.100 & 1.110 & 1.120 & 1.116 & 1.110 & 1.114 & 1.113 \\
    \ce{Be2}      & 2.330 & \mc{-}& 2.410 & \mc{-}& 2.350 & 2.251 & 2.278 & 2.250 & 2.250 \\
    \bottomrule
    \end{tabular}
    \label{tab:EquilibriumDistances}
\end{table*}
\begin{table*}[hbt!]
    \small
    \centering
    \caption{Bonding energy errors relative to FCI/ASCI reference in kcal/mol}
    \begin{tabular}{l|*{8}{S[table-format=3.1, table-column-width=1.1cm]}}
    \toprule
     & \mc{HF} & \mc{RPA} & \mc{CASCI} & \mc{NEVPT2} & \mc{BARE} & \mc{cRPA} & \mc{mRPA} & \mc{cRPAph} \\
    \otoprule
    \ce{H2} (HF)     &  124.7 &   68.4 &   -9.2 &   -0.9 &    3.3 &    9.5 &   -7.5 &   -8.4 \\
    \ce{H2} (PBE)    &  130.3 &    8.6 &  -20.4 &   -5.0 &   -0.5 &    2.1 &  -25.7 &  -25.7 \\
    \ce{H6} (4,4)             &  363.1 &  228.9 &  204.0 &  \mc{-}&  138.0 &  182.4 &  103.4 &   97.3 \\
    \ce{H6} (6,6)             &  363.1 &  228.9 &  -34.8 &   -4.5 &   -1.4 &   15.1 &  -36.6 &  -38.0 \\
    \ce{N2}                   &  536.8 &  191.4 &  -20.9 &    8.7 &  -14.9 &   -0.1 &  -94.2 &  -98.2 \\
    \ce{Be2}                  &  \mc{-}&   -2.3 &  \mc{-}&   -1.0 &    1.2 &    0.2 &    1.3 &    1.2 \\
\bottomrule
    \end{tabular}
    \label{tab:BondingEnergyErrors}
\end{table*}
Calculating the ground state energy of benzene in its equilibrium state is an illustrative example, but remains a single-reference problem. Therefore, we also present some bond dissociation curves where a single determinant generally dominates at equilibrium bond length, whereas upon dissociation, multiple determinants become important. This makes the dissociation of small molecules, for which we can perform highly accurate calculations (sometimes even using FCI), an important test. RPA has proven to describe bond dissociation qualitatively correctly, but needs improvement at medium bond lengths.\cite{Fuchs2005DescribingTheory,Henderson2010ThePerspective} Here we can test whether the combination of RPA and downfolding gives this improvement. Furthermore, we will also show dissociation curves obtained with the perturbative correction of NEVPT2\cite{Angeli2001IntroductionTheory} on top of a CASCI wave function, using the orbitals obtained from the preceding AMS calculation. We first consider a few covalently bonded molecules, \ce{H2}, \ce{H6}, and \ce{N2}, after which we will look at the much more difficult \ce{Be2}.\cite{Merritt2009BerylliumBonding} For \ce{H2}, \ce{H6}, and \ce{Be2} we limit all cRPA and cRPAph calculations to $\omega \rightarrow 0$. For $\ce{N2}$ we will go beyond this limit and explore the dependence on the screening frequency.

To properly describe the dissociation while using the partitioning of active space and environment, we use overlap tracking for active space selection. This is needed to avoid discontinuous dissociation curves due to inconsistent partitioning of active space and environment, caused by the changing nature of orbitals at different bond lengths. Therefore, we select the active space at the dissociation limit where the multi-reference character is most pronounced, necessitating an adequate active space the most. We then track the active space orbitals along the dissociation curve, by identifying at each bond length the orbitals that have the largest overlap with the active orbitals from the previous bond length. The overlap is evaluated through a dot product between orbital coefficient vectors in AO-basis.

\subsubsection{H\texorpdfstring{\textsubscript{2}}{2}}

\begin{figure}[hbt!]
    \centering
    \includegraphics[width=\linewidth]{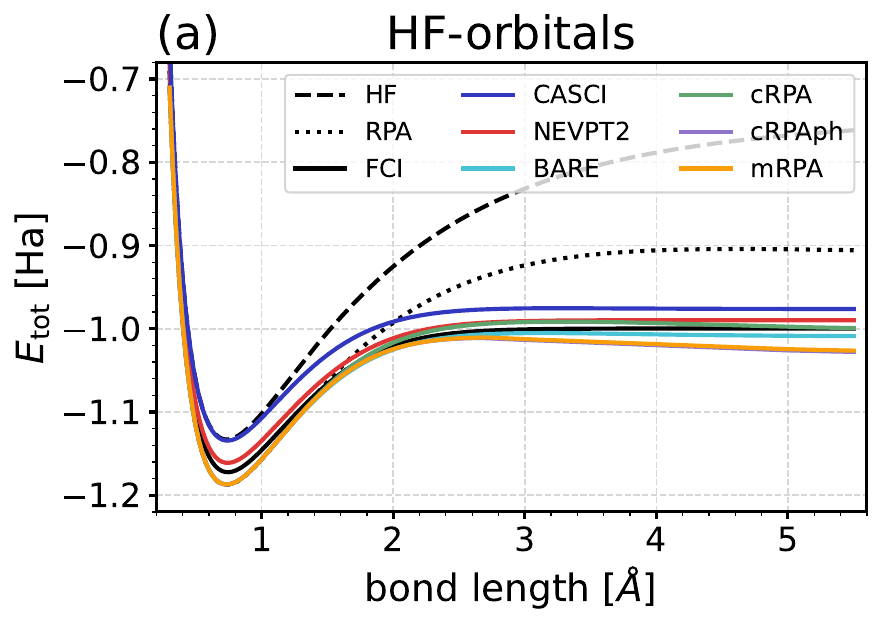}
    \includegraphics[width=\linewidth]{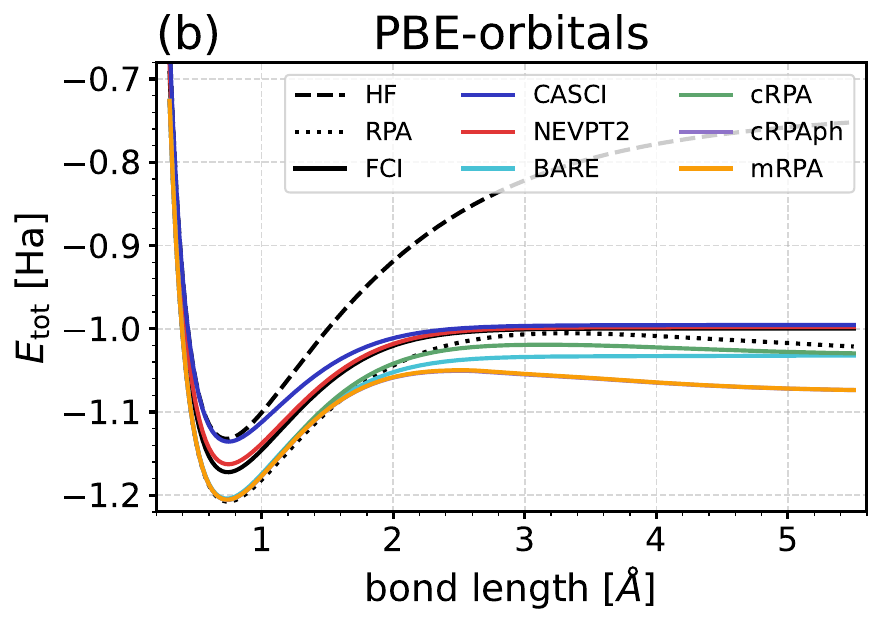}
    \caption{Dissociation curves of \ce{H2} in cc-pVTZ basis with (2,2) active space, composed of the bonding and anti-bonding linear combinations of the $1s$-orbitals. cRPA and cRPAph are evaluated at $\omega=0$. a) HF-orbitals, b) PBE-orbitals)}
    \label{fig:3_H2Dissociation}
\end{figure}
\begin{figure*}[hbt!]
    \centering
    \includegraphics[width=\linewidth]{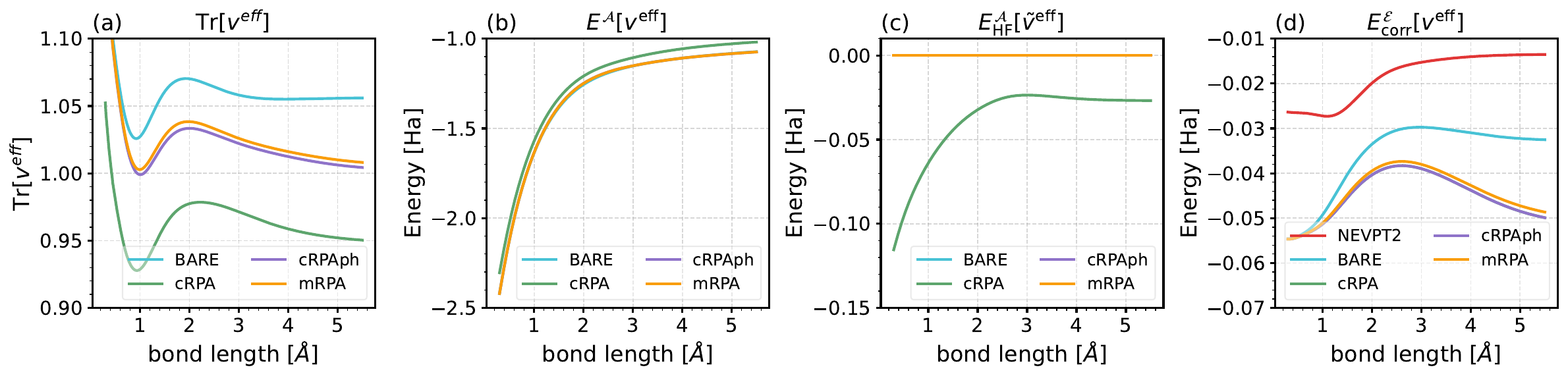}
    \caption{Trace of $v^\text{eff}$ and the energy contributions dependent on $v^\text{eff}$ along the dissociation curves for the methods with (2,2) active spaces of \ce{H2} in cc-pVTZ basis from a HF starting point. a) Trace of $v^\text{eff}$. b) Active space energy (BARE, mRPA, and cRPAph are almost identical and hence overlap). c) Mean-field correction accounting for the change in mean-field contribution from the active space due to using $v^\text{eff}$ instead of $v$ (strictly zero for BARE, mRPA, and cRPAph). d) Correlation energy of the environment. For the RPA-based methods, this is computed using Eq.~\eqref{eq:environment_correlation}, where cRPA and cRPAph give identical contributions. CASCI is excluded; all terms are equal to BARE, except for $E^\mE_\text{corr}$, which is zero for CASCI. For NEVPT2, only $E^\mE_\text{corr}$ is explicitly shown, all other terms are equal to BARE.}
    \label{fig:4_H2DissociationDecomposed}
\end{figure*}

Fig.~\ref{fig:3_H2Dissociation} shows the dissociation curves for \ce{H2} in the cc-pVTZ basis, using (Fig.~\ref{fig:3_H2Dissociation}a) HF or (Fig.~\ref{fig:3_H2Dissociation}b) PBE orbitals, both in the limit $\omega\rightarrow0$. Before going into detail, we first comment on the dependence on the mean-field orbitals. When we compare HF and PBE orbitals, the shape of the curves remains similar. The top two rows of Table ~\ref{tab:EquilibriumDistances} show that, both from a HF and PBE starting point, the equilibrium bond length is reproduced relatively well. However, the total energies are much more dependent on the orbitals. With HF orbitals, the methods that include the $E_\text{RPA,corr}^\mE$ contribution are in very close agreement with FCI, whereas with PBE orbitals they are too low across the whole dissociation curve. From the top two rows of Table~\ref{tab:BondingEnergyErrors}, there is also no clear pattern when comparing the starting points. From now on, we will only consider HF orbitals, keeping in mind the strong dependency on the mean-field orbitals.

There are some interesting subtleties when we compare the shape of the curves in Fig.~\ref{fig:3_H2Dissociation}a obtained using the different methods. At small atom distances, all RPA-based downfolding methods behave very similar to RPA. At medium atom distances, the system's multi-reference character increases, which the downfolded effective Hamiltonians start capturing. In this region, differences between screening methods can be observed, and screening the interactions leads to differences in energy compared to bare interactions. The RPA-based downfolding methods have a local maximum that they inherit from the RPA, consistent with the calculations by Henderson and Scuseria.\cite{Henderson2010ThePerspective} Additionally, if the bond is even more stretched, bare interactions lead to the expected behavior of converging to a dissociation energy, whereas the screened methods do not. While cRPA-screening compares well with FCI in the dissociation limit, mRPA and cRPAph overstabilize the system, and do not yet dissociate the bond correctly. 

Fig.~\ref{fig:4_H2DissociationDecomposed} scrutinizes the dependence of the individual energy components on $v^\text{eff}$, showing the trace and its constituent terms for the methods with an (2,2) active space. From Fig.~\ref{fig:4_H2DissociationDecomposed}a it becomes clear that mRPA and cRPAph screen the active space electrons almost identically, whereas cRPA screening is bigger in magnitude, consistent with the results described in Section~\ref{sec:BenzeneOmega}. Fig.~\ref{fig:4_H2DissociationDecomposed}b shows that the active space energy using cRPA is much higher compared to the ones obtained with a bare interaction. However, Fig.~\ref{fig:4_H2DissociationDecomposed}c shows that this is largely due to the smaller mean-field contribution, which is added back through the $E_\text{HF}^\mA[\Tilde{v}^\text{eff}]$ term in the environment energy. For mRPA and cRPAph, the screening has some appreciable effect on the active space energy only at medium bond lengths. This is partly because the mean-field contribution does not change compared to the one obtained with the bare interactions (also shown in Fig.~\ref{fig:4_H2DissociationDecomposed}c). The correlation part of the environment energy, shown in Fig.~\ref{fig:4_H2DissociationDecomposed}d, mostly explains the difference between mRPA/cRPAph and bare interactions. Upon dissociation, the difference from the $E^\mA_\text{RPA,corr}$ contribution (which is zero for $v^\text{eff}=v$) increases, explaining the unexpected behavior of no convergence. cRPA and cRPAph have identical contributions through $E_\text{RPA,corr}$, because only the particle-hole matrix elements enter this term. This also explains why cRPA is not yet completely dissociated at 6~\AA, but here the effect is slightly alleviated through the other contributions.

\subsubsection{H\texorpdfstring{\textsubscript{6}}{6}}

\begin{figure*}[hbt!]
    \centering
    \includegraphics[width=.48\linewidth]{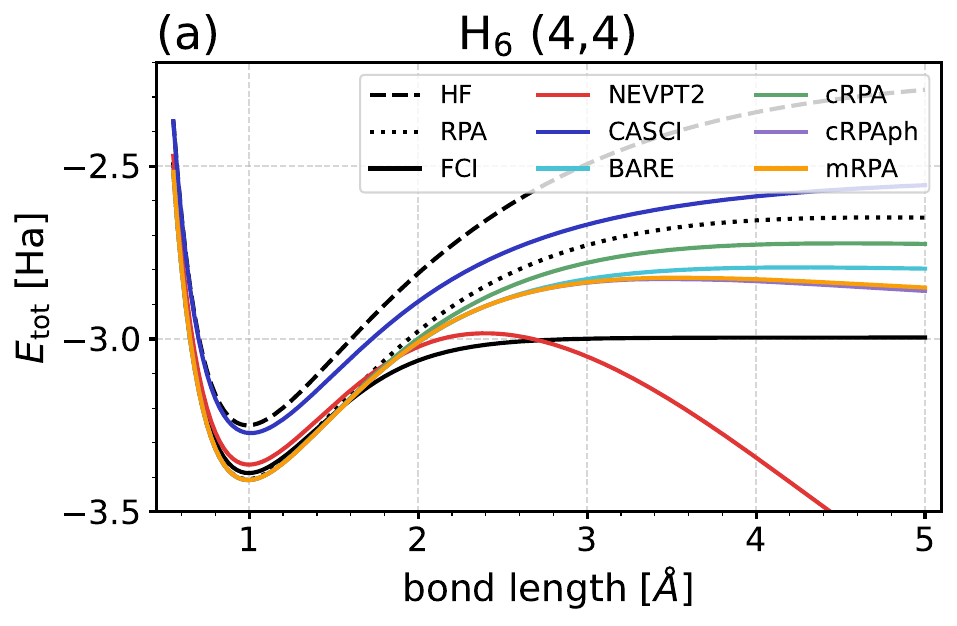}
    \includegraphics[width=.48\linewidth]{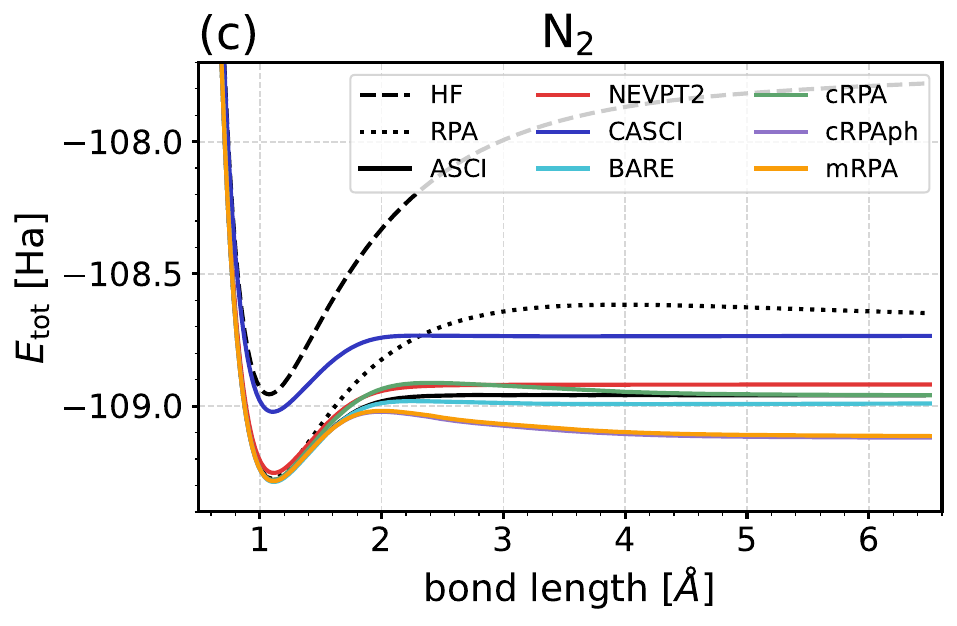}
    \includegraphics[width=.48\linewidth]{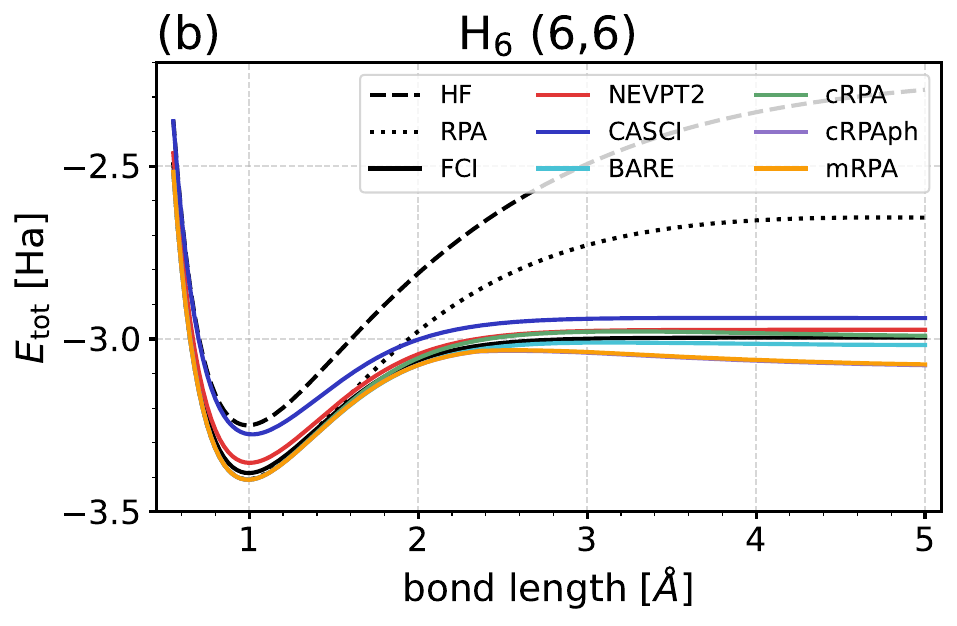}
    \includegraphics[width=.48\linewidth]{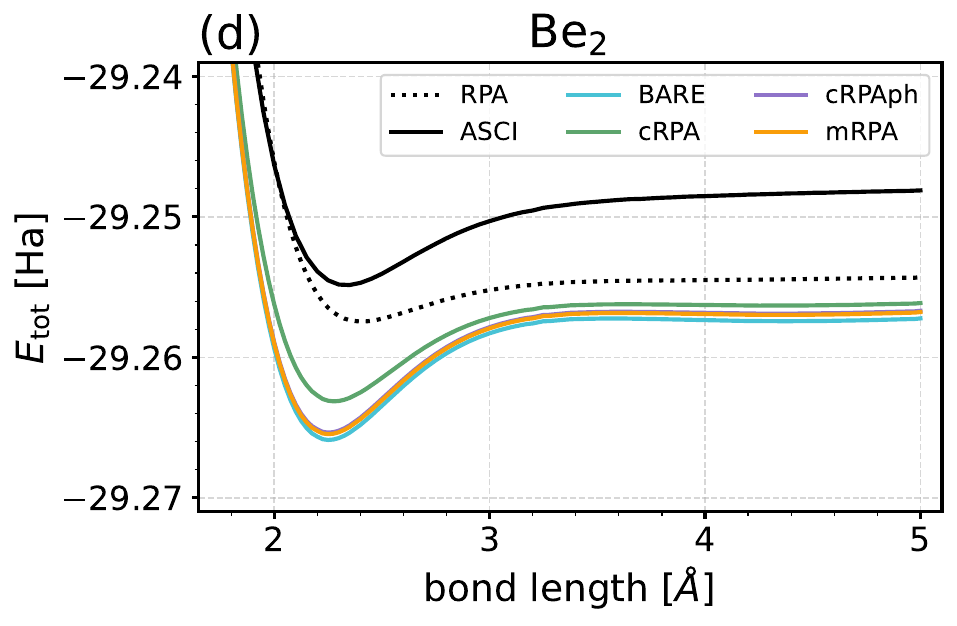}
    \caption{Dissociation curves of: a/b) \ce{H6} in cc-pVDZ basis with a) (4,4) active space b) (6,6) active space. Bond length is the distance between neighboring hydrogen atoms, which is equal for all six pairs. c) \ce{N2} in cc-pVDZ basis with a (6,6) active space, composed of all $2p$ linear combinations. The ASCI curve is obtained  with $N_\text{core}=32.000$ and $N_\text{target}=320.000$. d) \ce{Be2} in cc-pVTZ basis with a (4,4) active space, composed of $2s$ and $2p_z$ bonding and anti-bonding linear combinations. The ASCI curve is with $N_\text{core}=32.000$ and $N_\text{target}=320.000$.}
    \label{fig:5_DissociationCurves}
\end{figure*}

When using methods that can go beyond single reference, \ce{H2} is a fairly simple system, and a very small (2,2) active space is already sufficient to properly describe the dissociation. A small extension of the problem is the dissociation of a ring of six hydrogen atoms, \ce{H6}, which is also a well-studied system.\cite{Mejuto-Zaera2024QuantumAnsatz,Lan2015Communication:Chemistry} In the dissociation limit, the six linear combinations of $1s$ orbitals become degenerate, so at least a (6,6) active space is needed to describe the strong correlations. Fig.~\ref{fig:5_DissociationCurves}a/b show the results of the dissociation of \ce{H6}, with (4,4) and (6,6) active spaces, respectively. As expected, the (4,4) is not suitable for this system: the NEVPT2 energy diverges at larger bond lengths and the RPA-based methods drastically overestimate the bonding energies, as shown in Table~\ref{tab:BondingEnergyErrors}. With the (6,6) active space, the correlations are instead well-captured in this region. Other than this, the same conclusions as for \ce{H2} can be drawn, also when looking at the interpolated equilibrium bond lengths, shown in Table~\ref{tab:EquilibriumDistances}.

\subsubsection{N\texorpdfstring{\textsubscript{2}}{2}}

\begin{figure*}[hbt!]
    \centering
    \includegraphics[width=.48\linewidth]{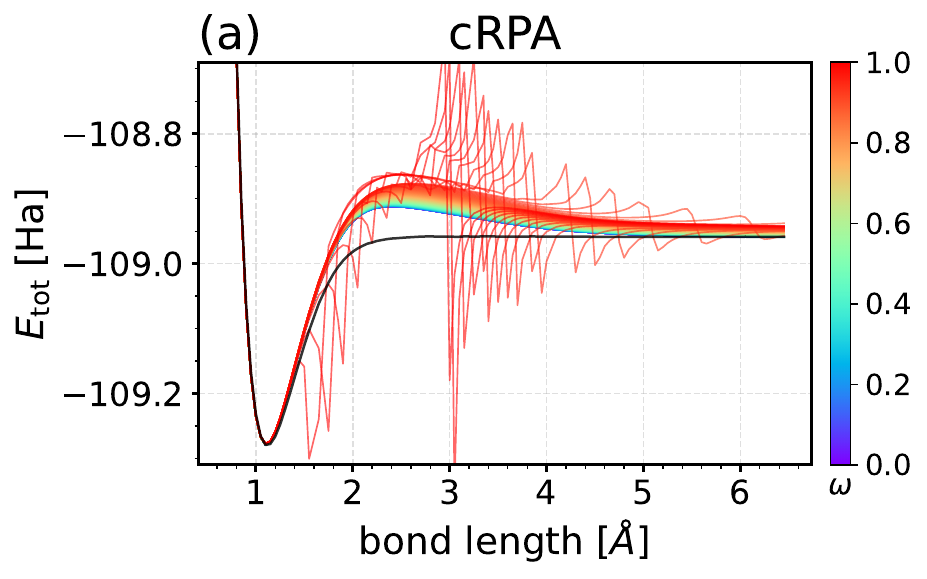}
    \includegraphics[width=.48\linewidth]{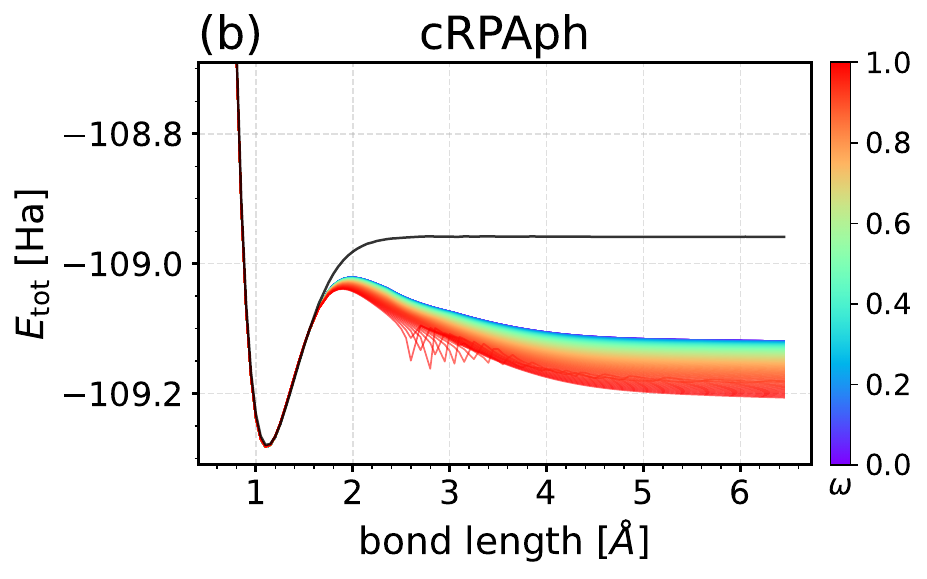}
    \caption{Dissociation curves of \ce{N2} in cc-pVDZ basis with a (6,6) active space, composed of all $2p$ linear combinations. The reference ASCI curve (black line) is obtained  with $N_\text{core}=32.000$ and $N_\text{target}=320.000$. a) Effect of $\omega$ on cRPA, with ASCI as reference. b) Effect of $\omega$ on cRPAph, with ASCI as reference. We used $\eta^+=10^{-3}$ as broadening factor.}
    \label{fig:S2_N2DissociationScreening}
\end{figure*}

\ce{N2} is also an interesting system to look at, due to the triple bond. Single-reference methods fail to describe the triple bond, and an (6,6) active space (consisting of bonding and anti-bonding combinations of $2p$-orbitals) is needed to describe both the $\sigma$-bond and the two $\pi$-bonds. When using an (6,6) active space, \ce{N2} is reduced to a problem with the same difficulty as \ce{H2}, the only difference being the consistency of the composition of the active space. For \ce{H2} this is straightforward and the HOMO and LUMO are always the active orbitals. For \ce{N2}, this is not the case, but orbital tracking solves this problem. From Fig.~\ref{fig:5_DissociationCurves}c (cc-pVDZ basis), the interpolated equilibrium bond length in Table~\ref{tab:EquilibriumDistances}, and bonding energy errors in Table~\ref{tab:BondingEnergyErrors}, the same conclusions as for \ce{H2} are drawn. It is noteworthy that cRPA describes the bonding energy correctly, even though it inherits a local maximum at medium bond length through the RPA description of the residual electron correlations outside the active space.\cite{Henderson2010ThePerspective} In the supplementary material, the results are also shown with a cc-pVTZ basis, and no qualitative differences can be observed.

So far, we have only considered a screening frequency of $\omega =0$, that exploits the separation of energy scales between the active space and its environment. A fixed value of $\omega \neq 0$ is difficult to justify physically, as it pushes the denominators towards the poles of the residual polarization. This is nicely illustrated in Fig.~\ref{fig:S2_N2DissociationScreening} that shows the dissociation curves for cRPA (a) and cRPAph (b) for a continuous range of screening frequencies. Most notable are divergences at higher screening frequencies, which is explained by the fact that there the constraint $\omega<\min\{\boldsymbol{\Omega}^\mR\}$ is not satisfied along the full dissociation curve. Around the equilibrium bond length, where a single determinant dominates, results are insensitive with respect to $\omega$. With increasing bond length and multi-reference character, the dependence on $\omega$ increases as well, and the dissociation curves are shifted away from their reference values. For cRPAph, this dependence strictly increases with the bond length. For cRPA the dependence decreases again at higher bond length.

\subsubsection{Be\texorpdfstring{\textsubscript{2}}{2}}

\ce{Be2} is not as straight-forward as \ce{N2}, due to the unclear nature of the bond. The combination of being weak but still relatively short is puzzling, and the shallow minimum has proven to be very sensitive to the basis set, level of theory, and active space size.\cite{Kaledin1999ElectronicResults, ElKhatib2014BerylliumCorrelation} Depending on these choices, sometimes the minimum is not even found,\cite{ElKhatib2014BerylliumCorrelation} suggesting that there is no bond, even though its existence has been proven experimentally.\cite{Kaledin1999ElectronicResults} This we can see in Table~\ref{tab:EquilibriumDistances}, which shows that CASCI does not have a minimum. The RPA-based downfolding methods yield more reasonable dissociation curves, as shown in Fig.~\ref{fig:5_DissociationCurves}d, with cRPA-screening leading to the closest agreement with the ASCI reference curve. In Table~\ref{tab:BondingEnergyErrors} we see that cRPA also predicts the bonding energy very well.

%% file: Images_new/1_BenzeneHeatmaps.tex
\begin{tikzpicture}
\pgfplotsset{
    colormap={cRPA}{
        rgb255(0cm)=(93,165,109);
        rgb255(1cm)=(255,255,255)
    },
    colormap={cRPAph}{
        rgb255(0cm)=(143,115,201);
        rgb255(1cm)=(255,255,255)
    },
    colormap={mRPA}{
        rgb255(0cm)=(249,157,9);
        rgb255(1cm)=(255,255,255)
    }
}
\begin{axis}[
    view={0}{90},
    colorbar,
    colorbar style={
        scaled y ticks = false,
        ytick={0,-0.01,-0.02,-0.03,-0.04},
        yticklabels={-0.00,-0.01,-0.02,-0.03,-0.04},
        width=0.3cm,
        tick label style={font=\small}
        },  
    colormap name = cRPA,
    point meta min=-0.04,
    point meta max=0.0,
    xtick=\empty,
    ytick=\empty,
    xmin=-0.5,
    xmax=20.5,
    ymin=-0.5,
    ymax=20.5,
    x dir=reverse,
    width=.315\textwidth,
    height=0.315\textwidth,
    axis equal image,
    title={(a)\qquad cRPA},
    title style={at={(-0.02,1.02)}, anchor=west}
]

\addplot[
    matrix plot*,
    point meta=explicit,
    mesh/ordering=rowwise,
    shader=flat corner,
    mesh/cols=21,
    draw=black!70,
    line width=0.25pt
] table [meta=C] {Data/cRPA.txt};

\end{axis}
\end{tikzpicture}
\begin{tikzpicture}
\pgfplotsset{
    colormap={cRPA}{
        rgb255(0cm)=(93,165,109);
        rgb255(1cm)=(255,255,255)
    },
    colormap={cRPAph}{
        rgb255(0cm)=(143,115,201);
        rgb255(1cm)=(255,255,255)
    },
    colormap={mRPA}{
        rgb255(0cm)=(249,157,9);
        rgb255(1cm)=(255,255,255)
    }
}
\begin{axis}[
    view={0}{90},
    colorbar,
    colorbar style={
        scaled y ticks = false,
        ytick={0,-0.01,-0.02,-0.03,-0.04},
        yticklabels={-0.00,-0.01,-0.02,-0.03,-0.04},
        width=0.3cm,
        tick label style={font=\small}
        },  
    colormap name = mRPA,
    point meta min=-0.04,
    point meta max=0.0,
    xtick=\empty,
    ytick=\empty,
    xmin=-0.5,
    xmax=20.5,
    ymin=-0.5,
    ymax=20.5,
    x dir=reverse,
    width=.315\textwidth,
    height=0.315\textwidth,
    axis equal image,
    title={(b)\qquad mRPA},
    title style={at={(-0.02,1.02)}, anchor=west}
]

\addplot[
    matrix plot*,
    point meta=explicit,
    mesh/ordering=rowwise,
    shader=flat corner,
    mesh/cols=21,
    draw=black!70,
    line width=0.25pt
] table [meta=C] {Data/mRPA.txt};

\end{axis}
\end{tikzpicture}
\begin{tikzpicture}
\pgfplotsset{
    colormap={cRPA}{
        rgb255(0cm)=(93,165,109);
        rgb255(1cm)=(255,255,255)
    },
    colormap={cRPAph}{
        rgb255(0cm)=(143,115,201);
        rgb255(1cm)=(255,255,255)
    },
    colormap={mRPA}{
        rgb255(0cm)=(249,157,9);
        rgb255(1cm)=(255,255,255)
    }
}
\begin{axis}[
    view={0}{90},
    colorbar,
    colorbar style={
        scaled y ticks = false,
        ytick={0,-0.01,-0.02,-0.03,-0.04},
        yticklabels={-0.00,-0.01,-0.02,-0.03,-0.04},
        width=0.3cm,
        tick label style={font=\small}
        },  
    colormap name = cRPAph,
    point meta min=-0.04,
    point meta max=0.0,
    xtick=\empty,
    ytick=\empty,
    xmin=-0.5,
    xmax=20.5,
    ymin=-0.5,
    ymax=20.5,
    x dir=reverse,
    width=.315\textwidth,
    height=0.315\textwidth,
    axis equal image,
    title={(c)\qquad cRPAph},
    title style={at={(-0.02,1.02)}, anchor=west}
]

\addplot[
    matrix plot*,
    point meta=explicit,
    mesh/ordering=rowwise,
    shader=flat corner,
    mesh/cols=21,
    draw=black!70,
    line width=0.25pt
] table [meta=C] {Data/cRPAph.txt};

\end{axis}
\end{tikzpicture}

%% file: Chapters/5_Conclusions.tex
\section{Conclusions}
We have presented an implementation of Green's function-based downfolding for molecules. We used RPA as low-level method to obtain a static effective Hamiltonian for the strongly correlated active space, while also providing a description of the dynamic correlation for the environment, avoiding much more demanding MRPT calculations. We did this by screening the environment following mRPA and cRPA (full and restricted to the particle-hole matrix elements), using the same constant screening frequency $\omega$ for all effective Hamiltonian terms. We compared the ground state energies of downfolded Hamiltonians obtained in this way for a few benchmark systems. We also investigated the effect of the choice of $\omega$, which is restricted by the neutral excitations in the residual space. The examples showed that the screened effective Hamiltonians yield energies that closely align with FCI or ASCI reference results. Total energies obtained with these interactions also improve over those obtained from mechanical embedding schemes where the active space Hamiltonian is constructed with bare interactions. In terms of predicting bonding energies, cRPA also performed remarkably well, on par with NEVPT2. Unlike cRPA, mRPA and cRPAph can yield too low energies in the dissociation limit due to a dominating dynamical correlation term. Interestingly, mRPA and the much simpler cRPAph screening essentially give indistinguishable screened interactions and dissociation curves in the static limit.

Overall, our results demonstrate that our relatively simple downfolding scheme has merit and should be tested for more interesting systems and properties. In this work, we targeted the total energy for an initial benchmark of these methods for molecules. However, effective Hamiltonian theories are arguably more suitable to describe how properties that are primarily determined by a small active region are tuned by their environment, such as excitation energies. Investigating such properties would also allow us to test whether the stronger screening we observed for cRPA relative to mRPA and cRPAph corresponds to over- or underscreening of the effective interactions between the active electrons. A concrete diagnostic would be to compare the active-space 2-particle RDM against that of a full-system FCI calculation projected on the active space. Furthermore, as motivated in the introduction, one of the primary use-cases for downfolded Hamiltonians will likely be large active spaces for which MRPT calculations are infeasible, and future work should target those.

To calculate RPA downfolded Hamiltonians for more complex systems, we are extending our implementation in multiple aspects. The implementation reported here solves the RPA equations using analytical integration.\cite{Bruneval2016MolgwClusters} This procedure scales as $\bigO(N^6)$ with system size, and is therefore not suitable for large systems. However, the proposed embedding scheme lays the foundation for the adoption of low-scaling, $\bigO(N^3)$, implementations of RPA,\cite{Kaltak2014LowTransformations, Wilhelm2018TowardAtoms, Vlcek2018SwiftCompression, Forster2020Low-OrderFitting, Duchemin2021Cubic-ScalingApproach, Spadetto2023TowardAccuracy, Yeh2023Low-ScalingSampling} which we will report on in a forthcoming publication. Furthermore, we will extend our code to spin-unrestricted calculations, and plan to improve over the simple mean-field orbitals by working in a natural orbital basis. Natural orbitals are a good substitute for the more costly optimization of MR wave functions\cite{Abrams2004NaturalWavefunctions} and can also guide active space selection.\cite{Jensen1988Second-orderCalculations} Making even more use of RPA, such natural orbitals are conveniently obtained from the RPA linearized density matrix.\cite{Bruneval2019ImprovedPotentials, Bruneval2019AssessmentMolecules, Ramberger2019RPAMethods}
Additional extensions can include extensions of the dynamic part of the self-energy through c$GW$\cite{Sheng2022GreensTheory, Romanova2023DynamicalStates}  and dynamical extensions, for instance by evaluating the 1-body and 2-body matrix elements of the effective Hamiltonian at specific frequencies evaluated through quasiparticle energies.\cite{Romanova2023DynamicalStates}

%% file: Chapters/appendix.tex
\appendix 
\section{\label{sec::appendix}Derivation of the effective Hamiltonian}
We here present a derivation of RPA downfolded Hamiltonians using the functional derivative approach.\cite{Biermann2003First-PrinciplesTheory} We start from the Baym--Kadanoff $\Phi$-functional in the $GW$ approximation\cite{Baym1961ConservationFunctions, Almbladh1999VariationalTheories}
\begin{equation}
    \label{eq:LW}
    \Phi_{GW} = \Phi_{\text{HF}} + \Phi_{GW,c}  \;,
\end{equation}
with 
\begin{equation}
  \Phi_{\text{HF}} = \frac{i}{2}\text{Tr}\left[\Sigma^{\text{HF}} G\right]
\end{equation}
and
\begin{equation}
    \Phi_{GW,c} = \frac{1}{2} \text{Tr} \left[\ln(1 - v P) + vP \right] \;.
\end{equation}
$P = -i G_sG_s$ is the RPA irreducible polarization and $G_s$ a non-interacting Green's function. Evaluated at $G_s$, $\Phi_c$ yields the electron correlation energy of the Klein energy functional\cite{Dahlen2006VariationalMolecules}, and $\Phi_{GW,c}[G_s] = E_{\text{RPA,corr}}$. 

By splitting the Green's function $G_s = G_s^{\mA} + G_s^{\mE}$, we can isolate the active space contribution to the polarizability
\begin{equation}
\label{eq:Psplit}
P = P^{\mA} + P^{\mR} \;,
\end{equation}
with 
\begin{equation}
P^\mR = -i(G_s^{\mA} G_s^{\mE} + G_s^{\mE}G_s^{\mA} + G_s^{\mE}G_s^{\mE}) \;.
\end{equation}
Our goal is to take a partial trace over the environment. While the HF functional is easily separated into an active-space contribution and a residual part, $\Phi_{\text{HF}} = \Phi^{\mA}_{\text{HF}} + \Phi^{\mR}_{\text{HF}}$, $\Phi_c$ can not be separated in this way since the logarithm is non-linear. Separating $\Phi_{GW,c}$ requires to define the effective interaction 
\begin{equation}\label{eq:A1veff}
v^{\text{eff}} = v + v P^{\mR} v^{\text{eff}} = (1 - vP^{\mR} )^{-1}v 
\end{equation}
through $P^\mR$. Substituting Eq.~\eqref{eq:Psplit} allows to rewrite the argument of the logarithm as\cite{Aryasetiawan2004Frequency-dependentCalculations},
\begin{equation}
\begin{aligned}
1 - v P = & 1 - vP^{\mR}  - vP^{\mA}  \\
= & 1-vP^{\mR} \\
 & - (1-vP^{\mR})(1-vP^{\mR})^{-1}vP^{\mA} \\
= & (1-vP^{\mR} )[1 - (1 - vP^{\mR})^{-1}vP^{\mA} ] \\
= & (1-vP^{\mR} )[1 - v^{\text{eff}}P^{\mA} ] \;,
\end{aligned}
\end{equation}
which allows us to separate the correlation part of the $\Phi$-functional as
\begin{equation}
    \label{eq:LW_split_c}
    \begin{aligned}
     \Phi_{GW,c}  = &
     \frac{1}{2}\text{Tr} \left[\ln(1 - v P^{\mR})+vP^{\mR}\right] \\
     + & 
     \frac{1}{2}\text{Tr} \left[\ln(1 - v^{\text{eff}} P^{\mA})\right] +
     \frac{1}{2}\text{Tr} [vP^{\mA}] \;.
     \end{aligned}
\end{equation}
To write down $\Phi^c_{GW}$ restricted to the active space, we must add and subtract $\frac{1}{2}\text{Tr} [v^{\text{eff}}P^{\mA}]$. We obtain
\begin{equation}
\label{eq:phi_c_split}
   \Phi_{GW,c} =  \Phi^{\mE}_{GW, c}[v] + \Phi^{\mA}_{GW, c}[v^\text{eff}] + \Phi^{\text{emb}}_{GW, c}[\tilde{v}^{\text{eff}}] \;,
\end{equation}
with the individual terms defined by 
\begin{align}
\label{eq:LW_r}
\Phi^{\mE}_{GW, c}[v] & =
      \frac{1}{2} \text{Tr} \left[\ln(1 - v P^\mR) + vP^\mR \right] \\
      \label{eq:LW_a}
  \Phi^{\mA}_{GW, c}[v^\text{eff}] & =    \frac{1}{2} \text{Tr} \left[\ln(1 - v^{\text{eff}} P^{\mA}) + v^{\text{eff}} P^{\mA} \right]  \\
   \label{eq:LW_emb}
  \Phi^{\text{emb}}_{GW, c}[\tilde{v}^{\text{eff}}] & = -
      \frac{1}{2}\text{Tr}\left[(v^{\text{eff}} - v)P^{\mA}\right] \\ \nonumber
      & = 
      -
      \frac{1}{2}\text{Tr}\left[\tilde{v}^{\text{eff}}P^{\mA}\right] \;.
\end{align}
Similarly, we rewrite the HF contribution to the $\Phi$-functional as
\begin{equation}
\label{eq:hf_split}
\begin{aligned}
    \Phi_{\text{HF}} & = \Phi^{\mE}_{\text{HF}}[v] + \Phi^{\mA}_{\text{HF}}[v] \\
    & = \Phi^{\mE}_{\text{HF}}[v] + \Phi^{\mA}_{\text{HF}}[v_{\text{eff}}] - 
    \Phi^{\text{emb}}_{\text{HF}}[\tilde{v}_{\text{eff}}] \;,
\end{aligned}
\end{equation}
where we have added and subtracted $\Phi^{\mA}_{\text{HF}}[v_{\text{eff}}]$ to arrive at the second equality. Together, Eq.~\eqref{eq:phi_c_split} and Eq.~\eqref{eq:hf_split} constitute an exact reformulation of Eq.~\eqref{eq:LW}. It is now convenient to define a residual $\Phi$-functional through 
\begin{equation}
     \label{eq:res_functional}
     \begin{aligned}
    \Phi^{\mR}[v] & =  \Phi[v] - \Phi^{\mA}[v^{\text{eff}}] \\
        & = 
        \Phi_{\text{HF}} - \Phi^{\mA}_{\text{HF}}[v^{\text{eff}}] + \Phi_{GW,c}
        - \Phi^{\mA}_{GW, c}[v^\text{eff}] \;.
    \end{aligned}
\end{equation}
We can make a static approximation for this residual $\Phi$-functional, for example by evaluating all matrix elements at $\omega=0$. Because $\Phi^{\mR}$ is defined for cRPA, it also holds for cRPAph, as it uses the same framework. For mRPA, we pragmatically assume the same form to hold.

\subsection{Effective 1-body term}\label{sec:A_eff1body}
To obtain the 1-body terms of the effective Hamiltonian, we take derivatives of Eq.~\eqref{eq:res_functional} with respect to $G^{\mA}$. This gives the matrix elements of the full system's self-energy in the active space, and allows us to define an embedding self-energy through
\begin{equation}
\label{eq:sigma_ebd}
\begin{aligned}
    \Sigma_{\text{emb}} = &
    \frac{\delta \Phi^{\mR}}{\delta G^{\mA}} \\
    = & 
    \frac{\delta}{\delta G^{\mA}}\left[\Phi_{\text{HF}} - \Phi^{\mA}_{\text{HF}}[v^{\text{eff}}] \right] \\
    & +
    \frac{\delta}{\delta G^{\mA}}\left[\Phi_{GW,c}
        - \Phi^{\mA}_{GW, c}[v^\text{eff}]\right] \;.
\end{aligned}
\end{equation}
To arrive at the effective Hamiltonian Eq.~\eqref{eq:H_eff_rpa}, we neglect the $GW$ contribution in $\Sigma^{\text{emb}}$. The HF embedding potential then becomes (making the non-trivial approximation that $\frac{\delta v^{\text{eff}}}{\delta G^{\mA}} = 0$)
\begin{equation}
\begin{aligned}
    \Sigma^{\text{HF}}_{\text{emb}} \approx & \frac{\delta \Phi^{\mR}_{\text{HF}}}{\delta G^{\mA}} \\
    = & 
    i\rho v - i \rho^{\mA}v^{\text{eff}} 
    + i G v - i G^{\mA}v^{\text{eff}} \\
    = & 
    i(\rho^\mE+\rho^\mA) v - i \rho^{\mA}v^{\text{eff}}\\
    & + i(G^\mE+G^\mA)v - i G^{\mA}v^{\text{eff}} \\
    = &
    i\rho^\mE+i\rho^\mA(v-v^\text{eff})\\
    & + iG^{\mE}v +iG^{\mA}(v-v^\text{eff})\\
    = & 
    i\rho^\mE v-i\rho^\mA\Tilde{v}^\text{eff}
    +iG^\mE{v}-iG^{\mA}\Tilde{v}^\text{eff} \\
    = & 
    f^{\mE}[v] - t^{\text{DC}} \;,
\end{aligned}
\end{equation}
with the double-counting correction defined in Eq.~\eqref{eq:tDC} and the Fock matrix contribution due to the environment with matrix elements
\begin{equation}
   f^{\mE}_{tu} = \sum^{\mE}_i (v_{tuii} - v_{tiiu})\rho_{ii} \;.
\end{equation}
Including the $GW$ contribution in $\Sigma^{\text{emb}}$ would give the additional terms
\begin{equation}
\begin{aligned}
    \frac{\delta \Phi^{\mR}_{\text{GW,c}}}{\delta G^{\mA}} = & iG(W-v)-iG^\mA(W-v^\text{eff})\\
    = & iG^\mE(W-v)+iG^\mA(W-v)\\
    & -iG^\mA(W-v^\text{eff})\\
    = & iG^\mE(W-v) + iG^\mA(v^\text{eff}-v)\\
    = & iG^\mE(W-v) + iG^\mA\Tilde{v}^\text{eff} \;,
\end{aligned}
\end{equation}
and  $\Sigma_\text{emb}$ would become
\begin{equation}
\begin{aligned}
    \Sigma_{\text{emb}} = & \frac{\delta \Phi^{\mR}_{\text{HF}}}{\delta G^{\mA}}+\frac{\delta \Phi^{\mR}_{\text{GW,c}}}{\delta G^{\mA}}\\
    = & i\rho^\mE v-i\rho^\mA\Tilde{v}^\text{eff}
    +iG^\mE{v}-iG^{\mA}\Tilde{v}^\text{eff}\\
    & + iG^\mE(W-v) + iG^\mA\Tilde{v}^\text{eff}\\
    = & i\rho^\mE{v}-i\rho^\mA\Tilde{v}^\text{eff}+iG^\mE W \;,
\end{aligned}    
\end{equation}
which is completely equivalent to the embedding self-energy derived in Ref.~\citenum{Sheng2022GreensTheory}.

\subsection{Constant energy shift}\label{sec:A_E0}
This form of double-counting correction has been referred to as "approximate" by Govoni and co-workers.\cite{Sheng2022GreensTheory} We stress, however, that this only refers to the neglect of the $GW$ contribution in the embedding self-energy. There is no double counting of any electron-electron interactions, as long as the electronic contribution $E^{\mE}$ to the constant energy shift $E_{0}$ of the effective Hamiltonian is defined appropriately, and $G_s$ is a HF Green's function. 

To derive the expression for this contribution we have used in our implementation (Eq.~\eqref{eq:environment_correlation}), we demand that the total energy of the system is exactly the sum of the scalar energy shift plus the residual energy produced by the solver $E_{\text{CAS}}$:
\begin{equation}
    E_{\text{tot}} = E^{\mE} + E_{\text{CAS}}
\end{equation}
The total energy of the system is given by\cite{Kotliar2006ElectronicTheory} 
\begin{equation}
    E_{\text{tot}} = \text{Tr}[h^{(0)}G^{\mE}] +  \text{Tr}[h^{(0)}G^{\mA}] + \Phi^{\mR} + \Phi^{\mA} \;.
\end{equation}
Since the residual energy contribution of the CAS solver is
\begin{equation}
   E_{\text{CAS}} = \text{Tr}[h^{(0)}G^{\mA}] +  \text{Tr}[\Sigma^{\text{HF}}_{\text{emb}}G^{\mA}]  + \Phi^{\mA} \;,
\end{equation}
we obtain
\begin{equation}
    E^{\mE} = \text{Tr}[h^{(0)}G^{\mE}] + \Phi^{\mR} - \text{Tr}[\Sigma^{\text{HF}}_{\text{emb}}G^{\mA}] \;.
\end{equation}
The last term in this expression evaluates to
\begin{equation}
\begin{aligned}
\text{Tr}[\Sigma^{\text{HF}}_{\text{emb}}G^{\mA}] = & (f - t^{\text{DC}}) G^{\mA} \\
= & 2E^{\mA-\mE}_{\text{HF}} - 2E^{\mA}_{\text{HF}}[\tilde{v}^{\text{eff}}] \;. 
\end{aligned}
\end{equation}
Combining this expression with $\Phi^{\mR}$ as defined in Eq.~\eqref{eq:res_functional}, The $\mA-\mE$ cross terms cancel exactly and we obtain
\begin{equation}
\begin{aligned}
    E^{\mE} = & E^\mE_\text{HF,core} + E^{\mE}_{\text{HF}} + E^{\mA}_{\text{HF}}[\tilde{v}^{\text{eff}}] \\
    &+ 
    E_{\text{RPA, corr}}[v] - E^{\mA}_{\text{RPA, corr}}[v^{\text{eff}}] \;,
\end{aligned}
\end{equation}
with
\begin{equation}
    E^\mE_\text{HF,core} = \text{Tr}[h^{(0)}G^{\mE}] \;.
\end{equation}

\section{mRPA}\label{sec:A_mRPA}
In Ref.~\citenum{Scott2023ATheory}, Scott, Backhouse and Booth derived relations between $\moment{n}$ and $\ApmB$. To make this work self-contained, we re-derive some of the relations for $\moment{0}$ and $\moment{1}$ here. The linear response Equation \eqref{eq:RPAcasida} can be expressed as
\begin{align}
    \ApB\XpY=&\XmY\boldsymbol{\Omega}\label{eq:CasidaRelation1}\\
    \AmB\XmY=&\XpY\boldsymbol{\Omega}\label{eq:CasidaRelation2}
\end{align}
Additionally, the normalization condition Eq.~\eqref{eq:XYnormalization} gives the relations
\begin{align}
    \XmY^T\XpY=\mathbb{1}\label{eq:NormCondition1}\\
    \XpY^T\XmY=\mathbb{1}\label{eq:NormCondition2}.
\end{align}

For $\moment{1}$, we can simply use Eqs.~\eqref{eq:CasidaRelation1} and \eqref{eq:NormCondition1}, so that
\begin{equation}
\label{eq:eta1_a}
\begin{aligned}
    \moment{1}=&\XpY\boldsymbol{\Omega}\XpY^T\\
    =&\AmB\XmY\XpY^T\\
    =&\AmB
\end{aligned}
\end{equation}
Additionally, by inserting the identity through Eq.~\eqref{eq:NormCondition1} and subsequently using Eq.~\eqref{eq:CasidaRelation1} in Eq.~\eqref{eq:eta1_a}, we can express $\moment{1}$ in terms of $\moment{0}$ (that is given by Eq.~\eqref{eq:eta_n}):
\begin{equation}
\begin{aligned}
    \moment{1}&=\XpY\boldsymbol{\Omega}\XpY^T\\
    &=\XpY\XpY^T\XmY\boldsymbol{\Omega}\XpY^T\\
    &=\moment{0}\XmY\boldsymbol{\Omega}\XpY^T\\
    &=\moment{0}\ApB\XpY\XpY^T\\
    &=\moment{0}\ApB\moment{0}.
\end{aligned}
\end{equation}
This also gives an expression for $\ApB$ in terms of $\moment{0}$ and $\moment{1}$, because multiplying from the left and from the right with $(\moment{0})^{-1}$ gives
\begin{equation}
    \ApB=(\moment{0})^{-1}\moment{1}(\moment{0})^{-1}
\end{equation}

Using Eqs.~\eqref{eq:A} and \eqref{eq:B} we obtain an expression for $v$ in terms of $\ApB$ and $\AmB$, for which we already derived an expression in terms of dd-moments;
\begin{equation}
\begin{aligned}
    v&=\frac{1}{2}[\ApB-\AmB]\\
    &=\frac{1}{2}[(\moment{0})^{-1}\moment{1}(\moment{0})^{-1}-\moment{1}].
\end{aligned}
\end{equation}

Lastly, the RPA correlation energy, which is commonly expressed following Ref.~\citenum{Furche2008DevelopingModel}:
\begin{equation}
\begin{aligned}
    E_\text{RPA,corr}&=\frac{1}{2}\text{Tr}[\boldsymbol{\Omega}-\tb{A}]\\
    &=\frac{1}{2}\left(\text{Tr}[\boldsymbol{\Omega}]-\text{Tr}[\tb{A}]\right),
\end{aligned}
\end{equation}
can also be rewritten in terms of dd-moments. For this, we manipulate Eq.\eqref{eq:CasidaRelation1} by multiplying from the right with $\XpY^T$, and subsequently use Eq.~\eqref{eq:NormCondition2} to get an expression for $\boldsymbol{\Omega}$:
\begin{equation}
    \boldsymbol{\Omega}=\XpY^T\ApB\XpY.
\end{equation}
Because we only need the trace of $\boldsymbol{\Omega}$, we can use the fact that the trace of a product of matrices is invariant under their cyclic permutations. Therefore,
\begin{equation}
\begin{aligned}
    \text{Tr}[\boldsymbol{\Omega}]&=\text{Tr}[\XpY^T\ApB\XpY]\\
    &=\text{Tr}[\XpY\XpY^T\ApB]\\
    &=\text{Tr}[\moment{0}\ApB],
\end{aligned}
\end{equation}
and
\begin{equation}\label{eq:mRPA_EcorrA}
    E_\text{RPA,corr}=\frac{1}{2}\text{Tr}[\moment{0}\ApB-\tb{A}] \;.
\end{equation}